\documentclass[aps,prr,twocolumn,longbibliography]{revtex4-2}
\usepackage[T1]{fontenc}
\usepackage[latin9]{inputenc}
\usepackage{verbatim}
\usepackage{float}
\usepackage{amsmath}
\usepackage{amssymb}
\usepackage{graphicx}
\usepackage{caption}
\usepackage{placeins}
\usepackage{capt-of}
\usepackage{ragged2e}
\usepackage{tikz}
\usetikzlibrary{backgrounds}
\usepackage[justification=justified,singlelinecheck=false]{caption}

\makeatletter

\usepackage{dblfloatfix}
\usepackage{float}
\usepackage{afterpage}
\ifdefined\showcaptionsetup
 \PassOptionsToPackage{caption=false}{subfig}
\fi
\usepackage{subfig}
\makeatother

\usepackage{babel}
\begin{document}
\title{Elastohydrodynamic instabilities of a soft robotic arm in a viscous
fluid}
\author{Mohamed Warda}
\email{mrmaw2@cam.ac.uk}

\affiliation{Cavendish Laboratory, University of Cambridge, J.J. Thomson Avenue,
Cambridge CB3 0US, United Kingdom}
\affiliation{Department of Applied Mathematics and Theoretical Physics, Centre
for Mathematical Sciences, University of Cambridge, Cambridge CB3
0WA, United Kingdom}
\author{Ronojoy Adhikari}
\affiliation{Department of Applied Mathematics and Theoretical Physics, Centre
for Mathematical Sciences, University of Cambridge, Cambridge CB3
0WA, United Kingdom}
\begin{abstract}
The design and control of soft robots operating in fluid environments
requires a careful understanding of the interplay between large elastic
body deformations and hydrodynamic forces. Here we show that this
interplay leads to novel elastohydrodynamic instabilities in a clamped
soft robotic arm driven terminally by a constant pressure in a viscous
fluid. We model the arm as a Cosserat rod that can stretch, shear
and bend. We obtain invariant, geometrically exact, non-linear equations
of motion by using Cartan's method of moving frames. Stability to
small perturbations of a straight rod is governed by a non-Hermitian
linear operator. Eigenanalysis shows that stability is lost through
a Hopf bifurcation with the increase of pressure above a first threshold.
A surprising return to stability is obtained with further increase
of pressure beyond a second threshold. Numerical solutions of the
non-linear equations, using a geometrically exact spectral method,
confirms stable limit-cycle oscillations between these two pressure
thresholds. An asymptotic analysis in the beam limit rationalizes
these results analytically. This counterintuitive sequence of bifurcations
underscores the subtle nature of the elastohydrodynamic coupling in
Cosserat rods and emphasizes their importance for the control of the
viscous dynamics of soft robots.
\end{abstract}
\maketitle

\section{Introduction}

The geometric dynamics of articulated arms has been studied extensively
due to their importance in robotics \citep{Murray1994,Selig2004,Chirikjian2009}.
In the emerging field of soft robotics, arms with continuously distributed
articulations have been constructed and studied with applications,
among others, to minimally invasive surgery \citep{Chirikjian1994,Ranzani2016}
and medical robotics \citep{Campisano2020,Campisano2021}. Since each
cross-section of such a continuous soft robotic arm has the full degrees
of freedom of a rigid body, the deformation of the arm must include
stretch, shear, bend and twist. The classical beam theories do not
include all of these degrees of freedom \citep{Love1892,Ericksen1957,Antman2005}
and more complete descriptions of a soft slender body then become
necessary. The special theory of Cosserat rods provides such a description
and has been used to model soft robotic arms \citep{Gazzola2018,Campisano2020,Boyer2021,Gazzola2021,Campisano2021,Chen2022,Gazzola2025}. 

When soft robotic arms operate in a fluid environment, the fluid-structure
interaction must be included in the balance of forces and torques.
Recent work has investigated this fluid-structure interaction \citep{Gazzola2025}
at finite Reynolds numbers. On the other hand, in many biological
settings the fluid-structure interaction is governed by slow viscous
flow. In fact, this approximation has been widely used to study the
dynamics of filaments in viscous fluids. Elastohydrodynamic instablities
of terminally driven filaments have recently been studied in this
approximation \citep{Laskar2017,DeCanio2017,Ling2018,Man2019,Fily2020,Clarke2024,Link2024,Schnitzer2025}.
These studies have revealed oscillatory flutter instabilities that
arise through Hopf bifurcations. They are related to non-conservative
problems in structural stability, for instance Beck\textquoteright s
column \citep{Beck1952,Bolotin1963,Ziegler1968,Carr1979,Chen1987,Koch2000,Wang2004}.
The Euler-Bernoulli model in \citep{DeCanio2017} has been further
developed and studied in subsequent work \citep{Ling2018,Man2019,Fily2020,Clarke2024,Schnitzer2025},
exploring both the linear instability mechanisms and the rich nonlinear
dynamics of filaments driven by follower forces. The effect of fluid
viscoelasticity on elastohydrodynamic instabilities has been studied
recently \citep{Link2024}.

While there is a vast literature on such elastohydrodynamic instabilities
of filaments in viscous fluids, the majority of existing work models
filaments as inextensible and unshearable slender structures. In contrast,
we study the dynamics of a soft robotic arm including the fluid-structure
interaction in the limit of slow viscous flow and the Cosserat degrees of freedom. This yields elastohydrodynamic
equations where the additional degrees of freedom provided by the
Cosserat theory are included. The forces and torques acting on the
rod due to motion in the ambient fluid are included, in the Stokes
approximation, as proportional to the local velocity and angular velocity
of the rod. Working in the overdamped limit, we derive the geometrized
equations of motion for the Cosserat rod in a coordinate-invariant
and geometrically exact manner in terms of a geometric field theory
\citep{Yan2025}. 

Using these equations of motion, we study the response of the soft
robotic arm to an externally applied terminal pressure, as might be
obtained when the arm makes contact with an underwater obstacle. The
linearization of the field theory reveals a subtle interplay between
the Cosserat degrees of freedom and leads to non-Hermitian equations
of motion. While the ``filament limit'' or ``beam limit'' of our
field theory shows excellent agreement with the results found in \citep{DeCanio2017,Clarke2024,Schnitzer2025},
deviations from the these limits reveal both qualitatively and quantitatively
distinct features of the dynamics. We find that incorporating shear
and angular dissipation into the model leads to a monotonic reduction
of the critical values of the follower force. Allowing for Cosserat
degrees of freedom, we find a loss of stability through a Hopf bifurcation
for values of pressure that are smaller compared to the case of inextensible
and unshearable filaments. More crucially, we find that the stretch
degree of freedom in rods that can sustain large compression results
in non-trivial alterations to the stability. In particular, we identify
regimes in parameter space where the rod loses stability through a
Hopf bifurcation and subsequently regains stability for larger values
of the pressure. Additionally, we identify regimes where the stretching
eliminates the Hopf bifurcation entirely and no loss of stability
is observed. To complement our linear stability analysis, we numerically
integrate the geometric field theory by means of a geometrically exact
numerical scheme that reveals stable limit cycle oscillations. This
corresponds to the sustained beating of the rod under applied pressure. 

The rest of the paper is organized as follows. In Section \ref{sec:fieldtheory}, we summarize
the geometric field theory formalism that we employ in this work,
which is presented in detail in \citep{Yan2025}. In Section \ref{sec:elastohydrodynamics},
we present our constitutive model for the viscous elastohydrodynamics
of a soft robotic arm and derive covariant equations of motion as
systems of nonlinear PDEs. In Section \ref{sec:pressuredriven} we impose boundary conditions
corresponding to a pressure applied to one terminus of the rod to
model a pressure-driven robotic arm, highlighting its role in injecting energy into the system. We linearize the field theory
in Section \ref{sec:linearization}, identifying the circulatory nature of the pressure-driven
model in the process, and write down the equations of motion in terms
of a non-Hermitian linear operator. We then present the results of
our linear stability analysis in Section \ref{sec:linearanalysis} and the results of our
nonlinear numerical simulations in Section \ref{sec:simulations}. In Section \ref{sec:euler}, we
connect our work to the existing literature on filaments by deriving
a ``partial beam limit'' that elucidates the counterintuitive spectrum
observed in Section \ref{sec:linearanalysis}. We conclude by summarizing and discussing
extensions of our study.

\section{Geometric Field Theory for Planar Cosserat rods \label{sec:fieldtheory}}

The configuration of a Cosserat rod at time $t$ is described by its
centerline $\boldsymbol{r}(u,t)$ and three orthonormal frame vectors
$\boldsymbol{e}_{1}(u,t),\boldsymbol{e}_{2}(u,t),\boldsymbol{e}_{3}(u,t)$,
rigidly attached to the cross-section at material parameter $u$.
We choose $\boldsymbol{e}_{1}$ to be normal to the cross-section
and $\boldsymbol{e}_{2}$ and $\boldsymbol{e}_{3}$ to span the plane
of the cross-section. We restrict ourselves in this work to configurations
in which the centerline remains in the plane and $\boldsymbol{e}_{3}$
is normal to the plane. Then, the configuration of the rod is given
by a $3\times3$ matrix-valued field 
\begin{equation}
\varphi(u,t)=\begin{bmatrix}1 & 0 & 0\\
\boldsymbol{r}(u,t) & \boldsymbol{e}_{1}(u,t) & \boldsymbol{e}_{2}(u,t)
\end{bmatrix},
\end{equation}
where each column vector has two components relative to a fixed frame.
The matrix is a representation of $SE(2)$, the Lie group of proper
rigid motions in the Euclidean plane. The spatial and temporal derivatives
of the configuration, when resolved in the moving frame provided by
$\boldsymbol{e}_{1}$ and $\boldsymbol{e}_{2}$, can be expressed
as 
\begin{equation}
\quad\partial_{t}\varphi=\varphi V,\quad\partial_{u}\varphi=\varphi E,\label{eq:kinematics}
\end{equation}
where $V$ and $E$ are matrices in the Lie algebra $\mathfrak{se}(2)$ \cite{Yan2025}.
This Lie algebra is three-dimensional and its elements can be expressed
in the canonical basis. The components of the velocity $V$ and deformation
$E$ in this canonical basis are $V_{\alpha}=(v_{1},v_{2},\Omega)$
and $E_{\alpha}=(h_{1},h_{2},\Pi)$. In terms of the centerline and
frame vectors we have
\begin{align}
v_{1} & =\boldsymbol{e}_{1}\cdot\partial_{t}\boldsymbol{r},\quad v_{2}=\boldsymbol{e}_{2}\cdot\partial_{t}\boldsymbol{r},\quad\Omega=\boldsymbol{e}_{2}\cdot\partial_{t}\boldsymbol{e}_{1},\nonumber \\
h_{1} & =\boldsymbol{e}_{1}\cdot\partial_{u}\boldsymbol{r},\quad h_{2}=\boldsymbol{e}_{2}\cdot\partial_{u}\boldsymbol{r},\quad\Pi=\boldsymbol{e}_{2}\cdot\partial_{u}\boldsymbol{e}_{1}.
\end{align}
These components, and therefore the matrices $V$ and $E$, are invariant
under rigid motions $\boldsymbol{r}\rightarrow\boldsymbol{R}\cdot\boldsymbol{r}+\boldsymbol{b}$
and $\boldsymbol{e}_{i}\rightarrow\boldsymbol{R}\cdot\boldsymbol{e}_{i}$
where $\boldsymbol{R}$ is a proper orthogonal matrix and $\boldsymbol{b}$
is an arbitrary vector in the plane. The element of arc of the centerline
is $ds=\sqrt{h_{1}^{2}+h_{2}^{2}}\,du$, $h_{1}$ and $h_{2}$ are
measures of shear of the cross-section and $\Pi$ is a measure of
the bending of the centerline. Given $E$ at any instant, the configuration of the rod can be recovered up to a rigid motion. Conversely, two configurations with identical values of $E$ can only differ by a rigid motion. 

The equality of mixed partial derivatives of the configuration, $\partial_{u}\partial_{t}\varphi=\partial_{t}\partial_{u}\varphi$,
immediately yields the compatibility condition
\begin{equation}
\partial_{t}E=\partial_{u}V+[E,V]\equiv\mathcal{D}V.\label{eq:integrability}
\end{equation}
Here $\mathcal{D}=\partial_{u}+\mathrm{ad}_{E}$ is a covariant derivative
that acts on elements of the Lie algebra. We list the appropriate
covariant derivatives, adjoint maps, and their duals, in Appendix \ref{sec:geometry}.

In the absence of inertia, we may write down the dynamics of the Cosserat
rod covariantly in terms of a generalized stress $\Sigma$ and generalized
body force density $j$ in the dual Lie algebra $\mathfrak{se}(2)^{*}$.
The components in the canonical dual basis are $\Sigma_{\alpha}=(F_{1},F_{2},M)$
and $j_{\alpha}=(f_{1},f_{2},m)$. The generalized balance law in
terms of these components is 
\begin{equation}
\mathcal{D}_{\alpha\beta}^{*}\Sigma_{\beta}+j_{\alpha}=0,\label{eq:balancelaw}
\end{equation}
where $\mathcal{D}^{\ast}=\partial_{u}+\text{ad}_{E}^{\ast}$ is a dual covariant
derivative that acts on elements of the dual Lie algebra \cite{Yan2025}. 

The principle of material indifference requires constitutive laws
to be invariant under rigid motions. This invariance is most easily
imposed by expressing the kinematic and dynamic equations in the moving
frame. Then, kinematic quantities take values in the Lie algebra $\mathfrak{se}(2)$
and dynamic quantities take values in the dual Lie algebra $\mathfrak{se}(2)^{*}$.
In this paper, we restrict ourselves to rheological constitutive laws
of the form $\Sigma_{\alpha}=\Sigma_{\alpha}(E_{\alpha})$ and dynamical
constitutive laws of the form $j_{\alpha}=j_{\alpha}(V_{\alpha})$. 

We note that Eq. (\ref{eq:balancelaw}) is covariant in the following
sense. For a constant $SE(2)$ group element $g$, the rigid transformation
$\varphi\to g\varphi$ leaves $V$ and $E$ invariant, which is easily
deduced from their definitions. It is then evident that specifying
constitutive laws $\Sigma_{\alpha}=\Sigma_{\alpha}(E_{\alpha})$ and
$j_{\alpha}=j_{\alpha}(V_{\alpha})$ leads to the invariance of Eq.
(\ref{eq:balancelaw}) under rigid motions.

\section{Viscous elastohydrodynamics \label{sec:elastohydrodynamics}}

We now make constitutive choices that are appropriate for a soft robotic
arm in a viscous fluid. We consider a Cosserat rod of rest length
$L$ and a stress-free configuration 
\begin{equation}
\bar{\boldsymbol{r}}(u)=(u,0),\quad\bar{\boldsymbol{e}}_{1}(u)=(1,0),\quad\bar{\boldsymbol{e}}_{2}(u)=(0,1).
\end{equation}
The components of the deformation in this configuration are $\bar{E}_{\alpha}=(1,0,0)$.
The strain of an arbitrary configuration with respect to this reference
configuration is $E-\bar{E}$ and this is invariant under rigid motions.
The linear constitutive equations for a linearly elastic Cosserat
rod subject to linear Stokes drag are
\begin{equation}
\Sigma_{\alpha}=K_{\alpha\beta}(E_{\beta}-\bar{{E}}_{\beta}),\quad j_{\alpha}=-\Gamma_{\alpha\beta}V_{\beta},\label{eq:constitutive}
\end{equation}
where $K_{\alpha\beta}$ and $\Gamma_{\alpha\beta}$ are symmetric
positive definite stiffness and friction tensors respectively. For
simplicity, we choose these to be diagonal, given in components by
$K_{\alpha\beta}=\mathrm{diag}(k_{1},k_{2},k_{3})$ and $\Gamma_{\alpha\beta}=(\gamma_{1},\gamma_{2},\gamma_{3})$. 

The components of the isotropic stiffness tensor may be related to the material and geometric properties of the rod \cite{Antman2005}: $\mathcal{E}$ (Young's modulus), $G$ (effective shear modulus), $A$ (cross-sectional area), and $I$ (second moment of area) via
\begin{equation}
k_1 = \mathcal{E}A,\quad k_2 = G A, \quad k_3 = \mathcal{E}I.\label{eq:material}
\end{equation}
The assumption of a constant isotropic friction tensor corresponds to the resistive force closure, i.e. the leading order local term of slender-body hydrodynamics for a slender rod. For Cosserat-type kinematics, the same asymptotic structure applies but with an additional local relation between angular velocity and distributed hydrodynamic torque. More complete nonlocal formulations for rotating/slender bodies are available (e.g. rotating-filament slender-body theory), and a slender-body theory for special Cosserat rods in Stokes flow has been derived recently \cite{Garg2023}. In the slender limit $a/L \ll 1$, where $a$ is the radius of the cross-section of the rod, nonlocal hydrodynamic interactions provide corrections beyond the local diagonal form; we neglect these higher-order terms to obtain a tractable analytical theory.

The rheological constitutive law admits an energy function 
\begin{align}
\begin{aligned}\mathcal{U}[\varphi] & =\end{aligned}
 & \frac{1}{2}\int_{0}^{L}\mathrm{d}u\,\,\langle K(E-\bar{E}),E-\bar{E}\rangle\nonumber \\
= & \frac{1}{2}\int_{0}^{L}\mathrm{d}u\,\,k_{1}(h_{1}-1)^{2}+k_{2}h_{2}^{2}+k_{3}\Pi^{2},
\end{align}
where $\langle\cdot\,,\,\cdot\rangle$ denotes the canonical pairing
between $\mathfrak{se}(2)$ and its dual. This encodes the energetic
penalty for a deformation from a reference state that is a planar
rod with no stretch and no shear ($h_{1}=1$, $h_{2}=0$), and no
bending ($\Pi=0$). 

Substituting the constitutive relationships in Eq. (\ref{eq:constitutive})
into the balance law in Eq. (\ref{eq:balancelaw}) gives an instantaneous
relationship between the velocities and deformations, 
\begin{equation}
\Gamma_{\alpha\beta}V_{\beta}=\mathcal{D}_{\alpha\beta}^{*}[K_{\beta\mu}(E_{\mu}-\bar{E}_{\mu})].\label{eq:dynamics}
\end{equation}
Since the deformations specify the shape of the rod and the velocities
specify its motion, this key relationship provides a rigidity-invariant
map between space of shapes and the space of motions. This map contains
the entire dynamic content of rod motion. 

The equation of motion for the configuration can be obtained explicitly
using canonical coordinates $(x,y,\theta)$ for $SE(2),$
\begin{equation}
\varphi(u,t)=\begin{bmatrix}1 & 0 & 0\\
x(u,t) & \cos\theta(u,t) & -\sin\theta(u,t)\\
y(u,t) & \sin\theta(u,t) & \cos\theta(u,t)
\end{bmatrix}.\label{eq:coordinatization}
\end{equation}
We then have two equivalent ways of formulating the equations of the
Cosserat rod. 

First, we may close the equation by eliminating $E$ completely and
writing three nonlinear time-dependent partial differential equations
for the coordinates of the rod by combining Eq. (\ref{eq:kinematics}),
Eq. (\ref{eq:dynamics}), and the coordinatization in Eq. (\ref{eq:coordinatization})
\begin{align}
\partial_{t}\varphi & =\varphi V,\quad\partial_{u}\varphi=\varphi E,\nonumber \\
\Gamma_{\alpha\beta}V_{\beta} & =\mathcal{D}_{\alpha\beta}^{*}[K_{\beta\mu}(E_{\mu}-\bar{E}_{\mu})].\label{eq:gebmodel}
\end{align}
This system requires one initial condition and two boundary conditions
for $\varphi$. We focus on clamped-forced boundary conditions in
this paper, of the general form
\begin{equation}
\varphi(0,t)=\mathbb{I},\quad\Sigma_{\alpha}(L,t)=P_{\alpha}(t),\label{eq:gebBCs}
\end{equation}
where a prescribed generalized stress $P_{\alpha}$ is enforced at the $u=L$ boundary, or tip of the
Cosserat rod. We stress that this forced boundary condition is a nonlinear
boundary condition on $\varphi$ upon application of Eq. (\ref{eq:constitutive})
and Eq. (\ref{eq:kinematics}). 

Second, we may impose the relation $\partial_{u}\varphi=\varphi E$
only at the initial time $t=0$ and promote the compatibility condition
in Eq. (\ref{eq:integrability}) to a time evolution equation in the
Lie algebra, which enforces $\partial_{u}\varphi=\varphi E$ for all
time \cite{KikuchiThesis, Kikuchi2023}. In this manner, we obtain six time-dependent partial differential
equations for the coordinates of the rod and the deformations
\begin{align}
\partial_{t}\varphi & =\varphi V,\quad\partial_{t}E_{\alpha}=\mathcal{D}_{\alpha\beta}V_{\beta},\nonumber \\
\Gamma_{\alpha\beta}V_{\beta} & =\mathcal{D}_{\alpha\beta}^{*}[K_{\beta\mu}(E_{\mu}-\bar{E}_{\mu})],\label{eq:igebmodel}
\end{align}
where the clamped-forced boundary conditions are now imposed on $E$
via
\begin{equation}
\mathcal{D}_{\alpha\beta}^{*}\Sigma_{\beta}(0,t)=0,\quad\Sigma_{\alpha}(L,t)=P_{\alpha}(t),\label{eq:igebBCs}
\end{equation}
and an application of the constitutive law. In particular, the previously
nonlinear (in $\varphi$) boundary condition $\Sigma_{\alpha}(L,t)=P_{\alpha}(t)$
is now a linear boundary condition in $E$, while the previously linear
(in $\varphi$) clamping boundary condition is now a nonlinear boundary
condition in $E$, obtained through setting $V(0,t)=0$. Applying
the balance law in Eq. (\ref{eq:dynamics}) leads to $\mathcal{D}_{\alpha\beta}^{*}\Sigma_{\beta}=0$,
a ``covariant Neumann'' boundary condition. We note that in this formulation,
the dynamical equation is time-independent and in the dual Lie algebra,
the time evolution of the deformation is in the Lie algebra and the
time evolution of the configuration is in the Lie group. Thus, the
dynamics is partitioned over the group, the algebra and its dual algebra.
We use this formulation below for numerical simulations. 

\section{Pressure-driven soft arm \label{sec:pressuredriven}}

We now obtain the appropriate boundary conditions for a soft robotic
arm described by the elastohydrodynamic equations derived in the previous
section. We assume that an external pressure acts normal to a terminal cross-section
and that the resultant force $\mathcal{F}$, the integral of the pressure over the cross-section, points along the $-\boldsymbol{e}_{1}$
axis. In our formulation, this implies that the boundary condition
at the pressurized end is 
\begin{equation}
\Sigma_{\alpha}(L,t)=(-\mathcal{F},0,0).
\end{equation}
This boundary condition is the planar Cosserat analog of the classical follower-load condition in Beck's column, where the applied force remains tangent to the deformed configuration. In contrast to active-filament models where follower forces arise from distributed internal motor activity, here the force is localized at the tip and externally imposed by pressure. Physically, the arm is subject to follower force $-\mathcal{F} \boldsymbol{e}_1$ that always remains aligned with the local normal to the cross-section, $\boldsymbol{e}_1$, at the tip. In the Euler-Bernoulli beam limit, as in Section \ref{sec:euler}, $\boldsymbol{e}_1$ gets locked to the local tangent of the underlying centerline, and our equations reduce exactly to the standard follower-force filament models studied in \cite{DeCanio2017, Clarke2024, Schnitzer2025}. We anticipate an interesting
feature of the dynamics from the following observation. Differentiating
the energy with respect to time gives
\begin{equation}
\begin{aligned}\frac{d\mathcal{U}}{dt}& =\int_{0}^{L}\mathrm{d}u\,\,\langle K(E-\bar{E}),\partial_{t}E\rangle\\
 & =\int_{0}^{L}\mathrm{d}u\,\,\langle\Sigma,\mathcal{D}V\rangle\\
 & =\langle\Sigma(L,t),V(L,t)\rangle-\int_{0}^{L}\mathrm{d}u\,\,\langle\mathcal{D}^{*}\Sigma,V\rangle\\
 & =\langle\Sigma(L,t),V(L,t)\rangle-\int_{0}^{L}\mathrm{d}u\,\,\langle\Gamma V,V\rangle\\
 & =-\mathcal{F}v_{1}(L,t)-\int_{0}^{L}\mathrm{d}u\,\,\langle\Gamma V,V\rangle,
\end{aligned}
\end{equation}
where we have used the compatibility condition, the covariant dynamics,
the properties of the covariant derivative, and the positive definiteness
of $\Gamma$. We see that in the absence of $\mathcal{F}$, this model
is entirely dissipative, as expected. On the other hand, the $-\mathcal{F}v_{1}$
term at the boundary may, depending on its sign, inject energy into
or withdraw energy from the rod. When the tip of the arm undergoes lateral motion while subject to the follower force, the force acquires a component along the velocity, performing net positive work over a cycle. This geometric phase-shift between force direction and velocity allows sustained energy injection even in an overdamped environment, in direct analogy with classical non-conservative follower-force problems such as Beck's column.

\section{Linearized Equations \label{sec:linearization}}

We first examine small-amplitude motions of the rod about the state
of static equilibrium. We label quantities corresponding to this equilibrium
with the suffix $0$. Imposing the static condition $V_{\alpha}^{(0)}=(0,0,0)$
on the velocity and using the dynamical law and the boundary conditions
yields 
\begin{equation}
\Sigma_{\alpha}^{(0)}=(-\mathcal{F},0,0),\quad E_{\alpha}^{(0)}=\left(\nu,0,0\right),
\end{equation}
where $\nu=1-\mathcal{F}/k_{1}$ is a compression due to the applied
pressure. The configuration corresponding to this state of deformation
is obtained by integration
\begin{equation}
x^{(0)}=\nu u,\quad y^{(0)}=0,\quad\theta^{(0)}=0,
\end{equation}
confirming that the rod remains undeflected and unsheared due to the
applied pressure. We now proceed with the linearization using the geometric procedure outlined in \citep{Yan2025}. The
result is 
\begin{align}
\Gamma_{\alpha\beta}\partial_{t}\xi_{\beta} & =\mathcal{D}_{\alpha\beta}^{*}[K_{\beta\mu}\mathcal{D}_{\mu\rho}\xi_{\rho}]+\left(\mathrm{ad}_{\mathcal{D}\xi}^{*}\right)_{\alpha\beta}\Sigma_{\beta}^{(0)}\nonumber \\
 & \equiv\mathcal{M}_{\alpha\beta}\xi_{\beta},\label{eq:linearization}
\end{align}
where $\mathcal{M}$ is the operator
\begin{equation}
\mathcal{M_{\alpha\beta}}=\begin{bmatrix}k_{1}\partial_{u}^{2} & 0 & 0\\
0 & k_{2}\partial_{u}^{2} & -(k_{2}\nu+\mathcal{F})\partial_{u}\\
0 & (k_{2}\nu+\mathcal{F})\partial_{u} & k_{3}\partial_{u}^{2}-\nu(k_{2}\nu+\mathcal{F})
\end{bmatrix}
\end{equation}
and, as outlined in \citep{Yan2025}, the covariant derivatives are
taken with respect to $E^{(0)}$. Note that the first order correction
to the deformations $\mathcal{D}_{\alpha\beta}\xi_{\beta}=E_{\alpha}^{(1)}$
has components
\begin{equation}
h_{1}^{(1)}=\partial_{u}x^{(1)},\quad h_{2}^{(1)}=\partial_{u}y^{(1)}-\nu\theta^{(1)},\quad\Pi^{(1)}=\partial_{u}\theta^{(1)}.
\end{equation}
The linearized boundary conditions are then the mixed boundary conditions
on $\xi$
\begin{equation}
\xi_{\alpha}(0,t)=(0,0,0),\quad\mathcal{D}_{\alpha\beta}\xi_{\beta}(L,t)=(0,0,0).\label{eq:linearBCs}
\end{equation}

We close this section by noting that the excess energy of the rod
to first order in perturbation expressed in terms of the configuration
is 
\begin{align}
\mathcal{U}_{\mathrm{lin}} & =\frac{1}{2}\int_{0}^{L}\mathrm{d}u\,\,\left\langle K\mathcal{D}\xi,\mathcal{D}\xi\right\rangle \nonumber \\
 & =\frac{1}{2}\int_{0}^{L}\mathrm{d}u\,\,k_{1}(h_{1}^{(1)})^{2}+k_{2}(h_{2}^{(1)})^{2}+k_{3}(\Pi^{(1)})^{2}.
\end{align}
Thus we may express the linearized equations as
\begin{equation}
\Gamma_{\alpha\beta}\partial_{t}\xi_{\beta}=-\frac{\delta\mathcal{U}_{\mathrm{lin}}}{\delta\xi_{\alpha}}+f_{\alpha},
\end{equation}
where
\[
f_{\alpha}=\mathcal{F}\left(0,-\Pi^{(1)},h_{2}^{(1)}\right),
\]
showing that the linearized dynamics is a gradient flow for zero pressure,
with a Timoshenko energy $\mathcal{U}_{\mathrm{lin}}$ as the Lyapunov functional. On the other
hand, the dynamics is circulatory for non-zero pressure, with the
Timoshenko energy no longer being a Lyapunov functional. The circulatory
terms couple the vertical deflection and the angle, with the pressure
acting as a source proportional to the bend, $\mathcal{F}\Pi^{(1)}$,
in the deflection equation and as a sink proportional to the shear,
$\mathcal{F}h_{2}^{(1)}$, in the angular equation. The circulatory
character of the dynamics anticipates the instabilities and limit
cycle oscillations we will uncover through further analyses below. The
quadratic dependence on the pressure anticipates the non-monotonic
variation of stability as the pressure is varied between the limits
$0\leq\mathcal{F}<k_{1}$. We now turn to establishing these results. 

\section{Linear stability analysis \label{sec:linearanalysis}}

To analyze the stability of the pre-stressed configuration, we expand
the perturbations in Eq. (\ref{eq:linearization}) as 
\begin{equation}
\xi_{\alpha}(u,t)=\mathrm{Re}\left(\sum_{n=1}^{\infty}(\Xi_{\alpha})_{n}(u)\exp(\lambda_{n}t)\right),
\end{equation}
where $(\Xi_{\alpha})_{n}=(X_{n},Y_{n},\Theta_{n})$. This leads to
the coupled generalized eigenvalue problem
\begin{equation}
\mathcal{M_{\alpha\beta}}\,\Xi_{\beta}=\lambda\Gamma_{\alpha\beta}\Xi_{\beta},
\end{equation}
with generalized eigenmodes $\Xi_{\alpha}$ associated with generalized
eigenvalues $\lambda_{n}$. Equivalently, we have the eigenvalue problem
\begin{equation}
\mathcal{L_{\alpha\beta}}\Xi_{\beta}=\lambda\Xi_{\alpha}
\end{equation}
for $\mathcal{L_{\alpha\beta}}=\Gamma_{\alpha\mu}^{-1}\mathcal{M_{\mu\beta}}$,
the generator of the dynamics. Furthermore, the eigenmodes $\Xi_{\alpha}$
satisfy the boundary conditions in Eq. (\ref{eq:linearBCs}). As shown
in Appendix \ref{sec:hermiticity}, for $\mathcal{F}=0$, $\mathcal{L_{\alpha\beta}}$
is Hermitian with respect to the $\Gamma-$inner product
\begin{equation}
(\psi,\xi)_{\Gamma}=\int_{0}^{L}\mathrm{d}u\,\,\langle\Gamma\psi,\xi\rangle,
\end{equation}
where $\psi_{\alpha}$, $\xi_{\alpha}:[0,L]\to\mathbb{C}^{3}$ are
vectors in a Hilbert space $\mathcal{H}$. Thus, for all $\psi_{\alpha},\xi_{\alpha}\in\mathcal{H}$,
$(\psi,\mathcal{L}\xi)_{\Gamma}=(\mathcal{L}\psi,\xi)_{\Gamma}$,
and $\mathcal{L}^{\dagger}=\mathcal{L}$. Moreover, it is also negative
definite, so $(\xi,\mathcal{L\xi})_{\Gamma}\leq0$ for all $\xi_{\alpha}\in\mathcal{H}$.
Hence, in the absence of a follower force, the eigenvalues $\lambda$
of $\mathcal{L_{\alpha\beta}}$ are strictly negative, leading to
stable dynamics. On the other hand, Hermiticity is lost for $\mathcal{F}\neq0$
and $\mathcal{L}^{\dagger}\neq\mathcal{L}$. This non-Hermiticity
of $\mathcal{L_{\alpha\beta}}$ allows us to anticipate oscillatory
and potentially unstable dynamics in the linear regime \cite{Ashida2020}.
\onecolumngrid
\begin{widetext}

\begin{figure}[H]
\subfloat[\label{fig:Hopf1}]{\includegraphics[scale=0.3]{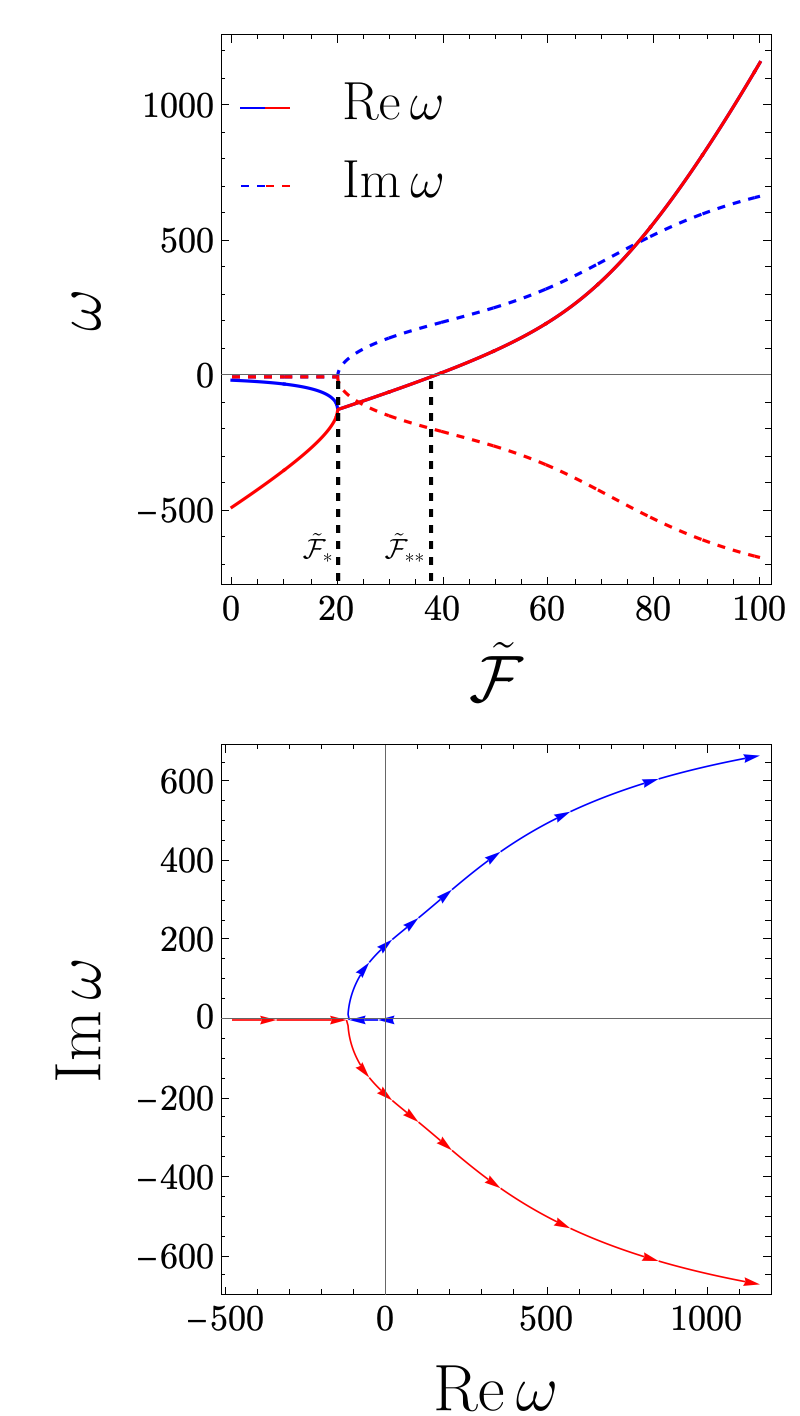}

}\subfloat[\label{fig:Hopf2}]{\includegraphics[scale=0.3]{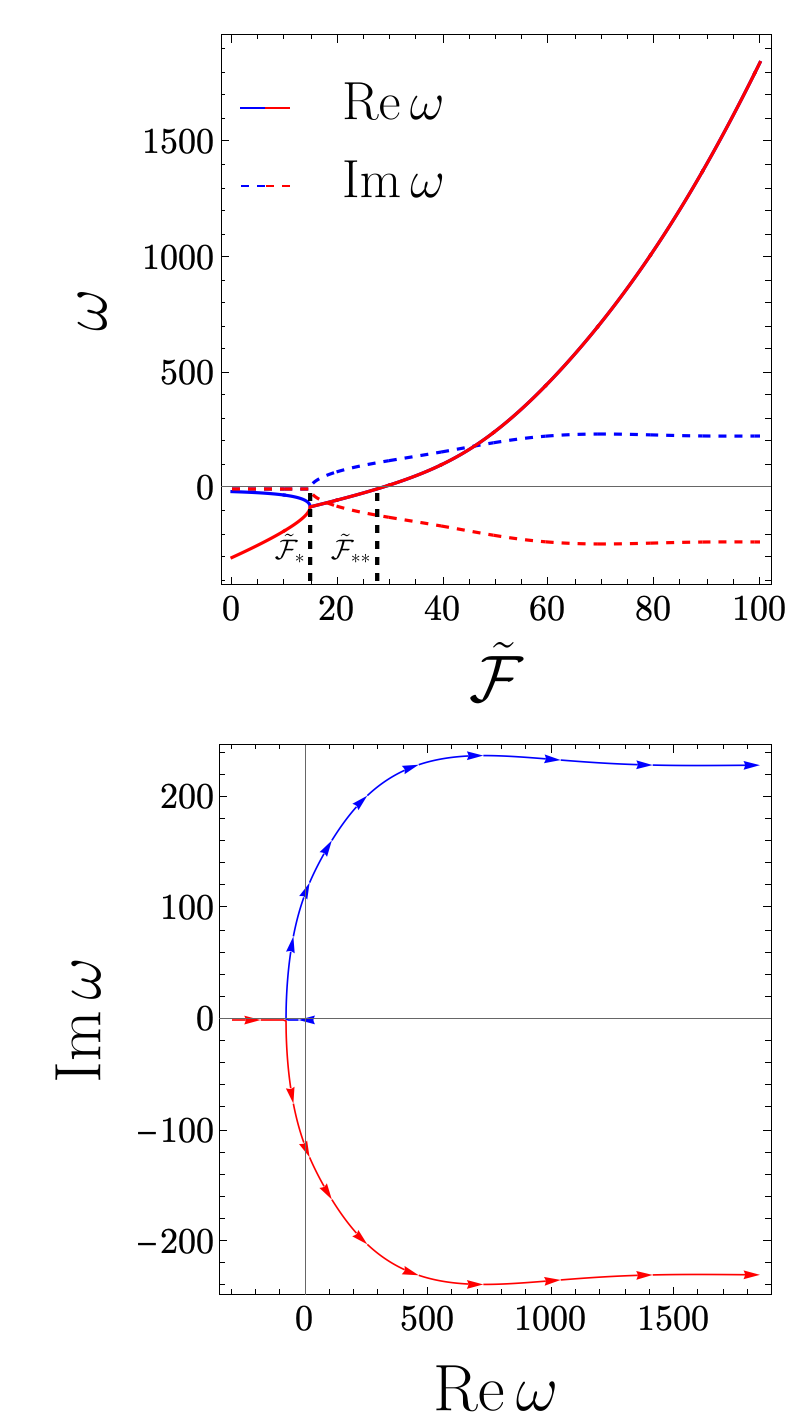}}\subfloat[\label{fig:Hopf3}]{\includegraphics[scale=0.3]{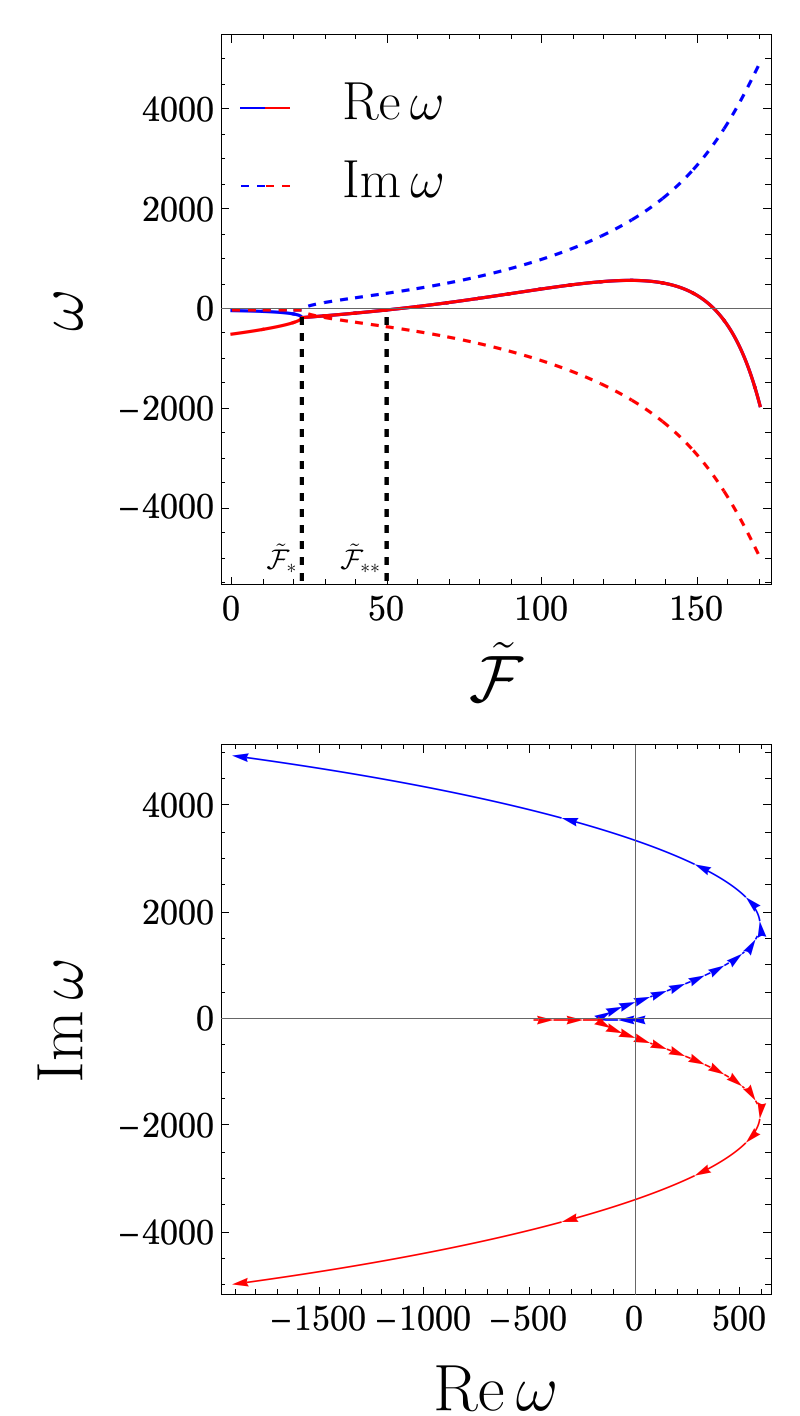}}\subfloat[\label{fig:Hopf4}]{\includegraphics[scale=0.3]{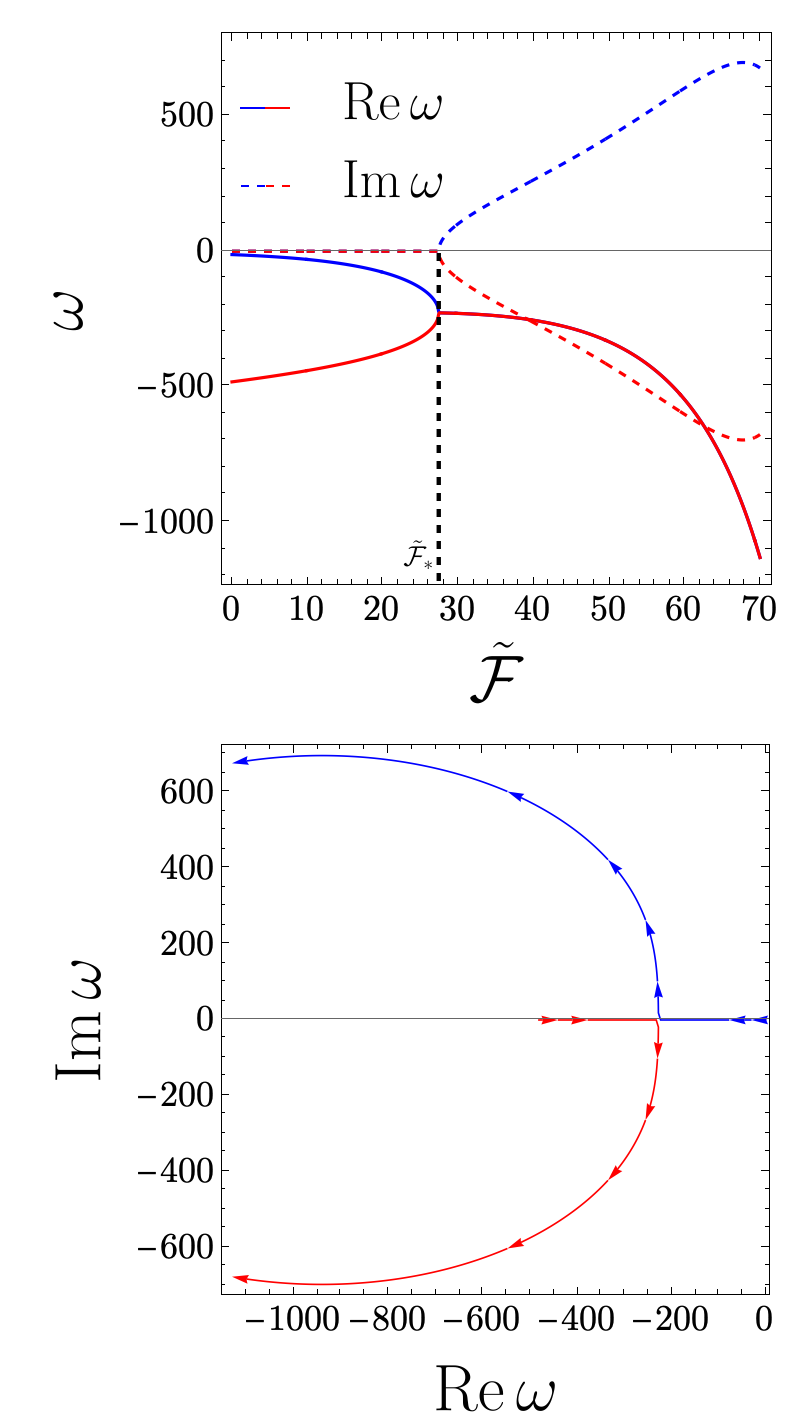}}\caption{The first two roots, $\omega_{1}$ (blue) and $\omega_{2}$ (red),
as a function of the strength of the follower force for the different
combinations of parameters considered in Table \ref{tab:parameters}.
The top plots show the real parts (solid lines) and the imaginary
parts (dotted lines) of $\omega_{1}$ and $\omega_{2}$. The corresponding
plots at the bottom display the locus of $\omega_{1}$ and $\omega_{2}$
as the follower force parameter is swept. In (a), We find an onset
of stable oscillations at $\tilde{\mathcal{F}}_*\approx20.01$, after which
$\omega_{1}$ and $\omega_{2}$ become complex conjugate pairs and
$\omega_{2}=\omega_{2}^{*}$. Stability is lost through a Hopf bifurcation,
which is observed for $\tilde{\mathcal{F}}_{**}\approx37.66$ when the real
parts of the roots become positive and the roots cross the imaginary
axis. A qualitatively similar behavior is observed in (b), where $\tilde{\mathcal{F}}_{*}\approx14.80$
and $\tilde{\mathcal{F}}_{**}\approx27.43$. In (c), we observe a qualitatively
different behavior, where stability is lost through a Hopf bifurcation
at $\tilde{\mathcal{F}}_{**}\approx49.76$ but subsequently regained for
$\tilde{\mathcal{F}}>150$ when the real parts of the roots become
negative again and the roots re-enter through the imaginary axis.
Finally, in (d), although we find an onset of stable oscillations
at $\tilde{\mathcal{F}}_{*}\approx27.41$, stability is not lost through
a Hopf bifurcation and no flutter instabilities are observed.}
\end{figure}

\end{widetext}
\twocolumngrid

It is evident from the block structure of $\mathcal{L_{\alpha\beta}}$
that the system decouples into a trivially stable diffusive longitudinal
component $X_{n}$ and a non-trivial transverse component coupling
$Y_{n}$ and $\Theta_{n}$. In other words, the system admits longitudinal
eigenmodes $(X_{n},0,0)$ with eigenvalues $\lambda_{n}^{\parallel}<0$,
and transverse eigenmodes $(0,Y_{n},\Theta_{n})$ with potentially
complex eigenvalues $\lambda_{n}^{\perp}$. In what follows, we will
adopt the notation $\omega_{n}\equiv\lambda_{n}^{\perp}$. The characteristic
polynomial for the transverse subsystem, when lengths are non-dimensionalized
by $L$, yields the dispersion relation
\begin{equation}
\begin{aligned} & \frac{\gamma_{3}}{k_{2}}\frac{\gamma_{2}L^{4}}{k_{3}}\omega^{2}\\
 & -\left[\left(\frac{\gamma_{2}L^{2}}{k_{2}}+\frac{\gamma_{3}L^{2}}{k_{3}}\right)q^{2}-\frac{\gamma_{2}L^{4}}{k_{3}}\nu\left(\nu+\frac{\mathcal{F}}{k_{2}}\right)\right]\omega\\
 & +q^{4}+\frac{\mathcal{F}L^{2}}{k_{3}}\left(\nu+\frac{\mathcal{F}}{k_{2}}\right)q^{2}=0,
\end{aligned}
\end{equation}
from which we identify the shortest time scale $\tau_{1}$ and the
longest time scale $\tau_{2}$ associated with the problem
\begin{equation}
\tau_{1}=\frac{\gamma_{3}}{k_{2}},\quad\tau_{2}=\frac{\gamma_{2}L^{4}}{k_{3}}.\label{eq:timescales}
\end{equation}
Any choice of non-dimensionalization yields dimensionless stiffness
and friction tensors $\tilde{K}_{\alpha\beta}$, $\tilde{\Gamma}_{\alpha\beta}$.
For convenience, we adopt the time scale $\tau_{1}$ for the nonlinear
simulations, and reserve $\tau_{2}$ for the linear stability analysis
we will describe below. We denote non-dimensionalized constitutive
parameters as in the components of $\tilde{K}$, $\tilde{\Gamma}$
by tildes, but we omit this notation for the degrees of freedom $(x,y,\theta)$
for clarity of notation. 
\begin{table}[t]
  \centering
  \begin{ruledtabular}
    \begin{tabular}{ccc}
      Symbol & Definition & Interpretation  \\
      \colrule
      $\tilde{k}_{1}=k_{1}L^{2}/k_{3}$ & Stretch stiffness ratio       & Compressibility       \\
      $\tilde{k}_{2}=k_{2}L^{2}/k_{3}$ & Shear stiffness ratio        & Shear rigidity        \\
      $\tilde{\gamma}_{3}=\gamma_{3}/{\gamma_{2}L^{2}}$ & Rotational drag ratio& Angular dissipation       \\
      $\tilde{\mathcal{F}}=\mathcal{F}L^{2}/{k_{3}}$ & Follower force strength        & Control parameter     \\
    \end{tabular}
  \end{ruledtabular}
 \caption{\parbox{\linewidth}{\justifying\noindent \newline A summary of the dimensionless parameters $(\tilde{k}_{1},\tilde{k}_{2},\tilde{\gamma}_{3}, \tilde{\mathcal{F}})$ introduced in this section and their interpretations.}\label{tab:summary}}
\end{table}
Adopting the time scale $\tau_{2}$ leads
to the non-dimensionalized parameters
\begin{equation}
\tilde{k}_{1}=\frac{k_{1}L^{2}}{k_{3}},\quad\tilde{k}_{2}=\frac{k_{2}L^{2}}{k_{3}},\quad\tilde{k}_{3}=1\label{eq:stiffness}
\end{equation}
for the stiffness,
\begin{equation}
\tilde{\gamma}_{1}=\frac{\gamma_{1}}{\gamma_{2}},\quad\tilde{\gamma}_{2}=1,\quad\tilde{\gamma}_{3}=\frac{\gamma_{3}}{\gamma_{2}L^{2}}\label{eq:friction}
\end{equation}
for the friction, and
\begin{equation}
\tilde{\mathcal{F}}=\frac{\mathcal{F}L^{2}}{k_{3}}\label{eq:bifurcationparameter}
\end{equation}
for the follower force. We summarize these definitions in Table \ref{tab:summary}. The constitutive parameters in Table \ref{tab:summary} are dimensionless stiffness ratios obtained by scaling forces with the bending scale $k_3/ L^2 = \mathcal{E}I/L^2$. In particular, using Eq. (\ref{eq:material}), $\tilde{k}_1 = A L^2/I$, which depends only on geometry (slenderness) and typically scales like $\tilde{k}_1 \sim (L/a)^2$. Thus values such as $\tilde{k}_1 \sim 10^2$ correspond to moderately slender beams, while truly slender rods $(L/a\gg 1)$ naturally have $\tilde{k}_1 \gg 10^2$. Likewise $\tilde{k}_2 = GA L^2/\mathcal{E}I$ and the ratio $\tilde{k}_2/ \tilde{k}_1 = G/\mathcal{E}$. For isotropic elastic materials, $G/\mathcal{E} = 1/[2(1 + \mu)]$, where $\mu$ is the Poisson ratio. Typically, for soft materials, $\mu \approx 1$ and $\tilde{k}_2/ \tilde{k}_1 \approx 1/3$ \cite{Jastrzebski1959}. Therefore, regimes with $\tilde{k}_2 \gg \tilde{k}_1$ should be interpreted as enforcing a near-unshearable limit or as requiring strongly anisotropic/engineered materials.

An application of the boundary conditions on $(Y,\Theta)$
as in Appendix \ref{sec:eigenvalues} allows us to locate the eigenvalues $\omega_{n}$
of the system through the transcendental equation
\begin{equation}
B_{1}+B_{2}\sinh\chi_{1}\sin\chi_{2}+B_{3}\cosh\chi_{1}\cos\chi_{2}=0,
\end{equation}
where $B_{1},B_{2},B_{3},\chi_{1},\chi_{2}$ are functions of $\omega$, $\tilde{k}_{1}$, $\tilde{k}_{2}$, $\tilde{\gamma}_{3}$, and $\tilde{\mathcal{F}}$, and are obtained explicitly in Appendix
\ref{sec:eigenvalues}.

To investigate the spectrum of $\mathcal{L}$, we fix a particular
physical regime of the Cosserat rod by specifying a tuple of constitutive
parameters $(\tilde{k}_{1},\tilde{k}_{2},\tilde{\gamma}_{3})$, and carrying out a sweep of $\tilde{\mathcal{F}}$
to control the strength of the follower force while locating the roots
of the transcendental equation. We show the variation of the two smallest
roots (in magnitude) $\omega_{1}$, $\omega_{2}$ of the transcendental
equation. In particular, we consider the physical parameters $(\tilde{k}_{1},\tilde{k}_{2},\tilde{\gamma}_{3})$
in Table \ref{tab:parameters}.

\begin{table}[t]
  \centering
  \begin{ruledtabular}
    \begin{tabular}{ccccc}
      Figure & $\tilde{k}_{1}$ & $\tilde{k}_{2}$ & $\tilde{\gamma}_{3}$ & Description \\
      \colrule
      1 & $10^{4}$        & $10^{4}$        & $10^{-4}$ & Inextensible, unshearable \\
      2 & $10^{4}$        & $10^{2}$        & $10^{-2}$ & Shearable \\
      3 & $2\times 10^{2}$& $10^{4}$        & $10^{-4}$ & Extensible \\
      4 & $10^{2}$        & $10^{4}$        & $10^{-4}$ & Extensible \\
    \end{tabular}
  \end{ruledtabular}
  \caption{\parbox{\linewidth}{\justifying\noindent The range of constitutive parameters $(\tilde{k}_{1},\tilde{k}_{2},\tilde{\gamma}_{3})$
corresponding to Fig. 2. Control over the elastic moduli allows us
to investigate the effects of planar Cosserat degrees of freedom (stretch
and shear).}\label{tab:parameters}}
\end{table}
The range of parameters $(\tilde{k}_{1},\tilde{k}_{2},\tilde{\gamma}_{3})$ is chosen to investigate the effect of the
Cosserat degrees of freedom on the spectrum, and the parameters $(\tilde{k}_{1},\tilde{k}_{2},\tilde{\gamma}_{3})=(10^{4},10^{4},10^{-4})$
precisely recover the Euler-Bernoulli beam limit of a Cosserat rod,
where the frame of directors is locked to the Frenet-Serret frame
of the underlying centerline. This limit is discussed in detail in
Section \ref{sec:euler}.

\begin{figure}[t]
\centering
\includegraphics[scale=0.4]{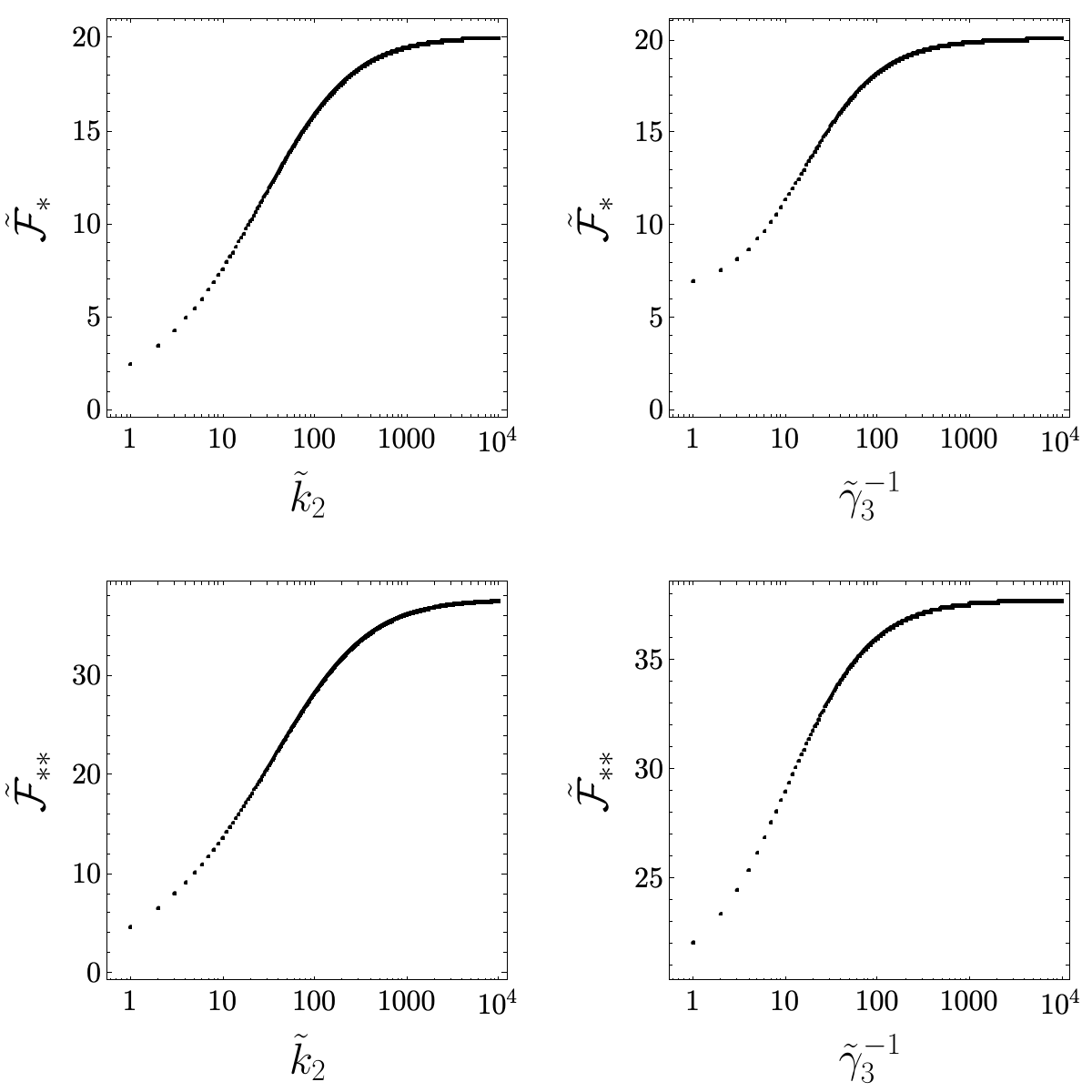}\caption{\label{fig:shearrelaxation}Semilog plots of the critical values of
the follower force strength as a function of $\tilde{k}_{2}$ and$\tilde{\gamma}_{3}^{-1}$.}
\end{figure}

We first investigate the spectrum for $\tilde{k}_{1}=10^{4}$. As
shown in Fig. \ref{fig:Hopf1}, for $\tilde{k}_{2}=10^{4}$, $\tilde{\gamma}_{3}=10^{-4}$,
we recover the classic flutter instability Hopf bifurcation diagram
observed in the overdamped dynamics of planar inextensible filaments
as in \citep{DeCanio2017}. Namely, we find an onset of stable oscillations
at a critical value of the follower force $\tilde{\mathcal{F}}\approx20.01$,
followed by a Hopf bifurcation/flutter instability at a second critical
value $\tilde{\mathcal{F}}\approx37.66$. On the other hand, as shown in
Fig. \ref{fig:Hopf2}, allowing shear and angular dissipation through
setting $\tilde{k}_{2}=10^{2}$, $\tilde{\gamma}_{3}=10^{-2}$ also
recovers a qualitatively similar but quantitatively different bifurcation
pattern. In particular, we now find that the onset of oscillations and Hopf bifurcation
occur for $\tilde{\mathcal{F}}\approx14.80$ and $\tilde{\mathcal{F}}\approx27.43$, respectively.
For completeness, we also compute and illustrate the transverse eigenmodes
and their associated deformations for different values of $\tilde{\mathcal{F}}$
in Fig. \ref{fig:eigenmodes10} -- \ref{fig:eigenmodes30} in Appendix
\ref{sec:eigenvalues}. To understand the impact of shear and angular dissipation on the
critical values of the follower force, we carry out a sweep of $\tilde{k}_{2}$
and $\tilde{\gamma}_{3}$. As shown in Fig. \ref{fig:shearrelaxation},
We find that both critical values of $\mathcal{F}$ increase monotonically
as a function of $\tilde{k}_{2}$ and $1/\tilde{\gamma}_{3}$, plateauing
at their Euler-Bernoulli values for sufficiently large $\tilde{k}_{2}$
and sufficiently small $\tilde{\gamma}_{3}$.

\begin{figure}[t]
\centering
\includegraphics[scale=0.47]{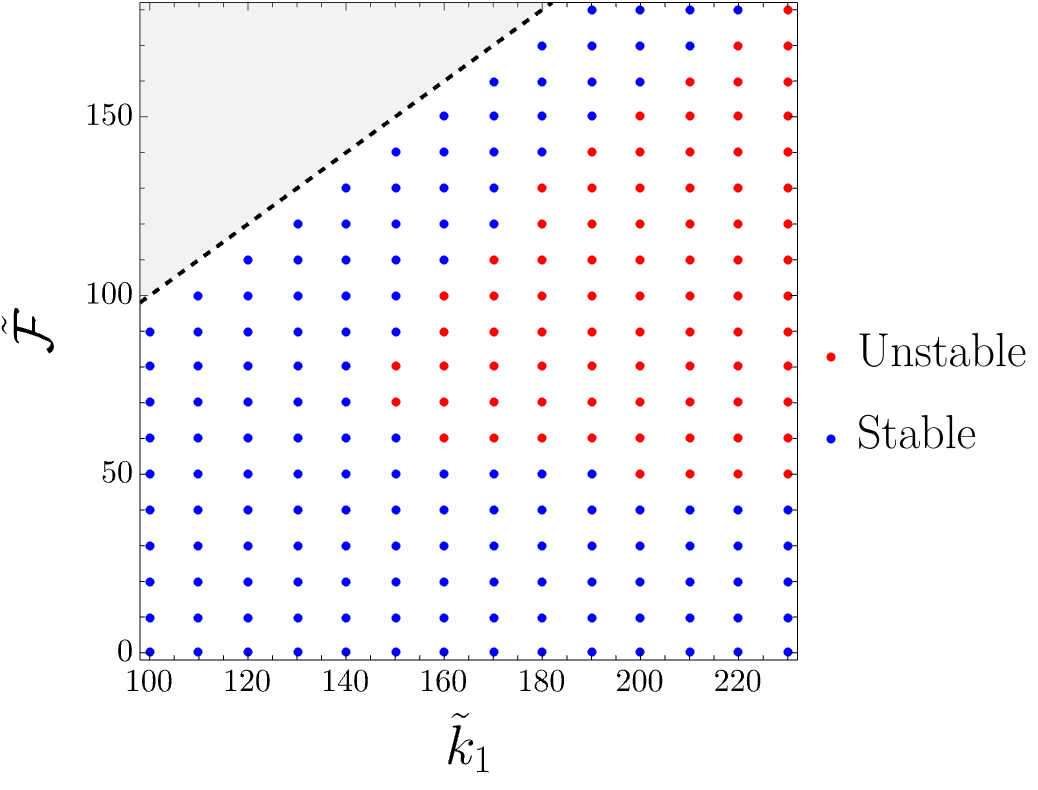}\caption{\label{fig:phase_schematic}Phase diagram in the $(\tilde {\mathcal{F}},\tilde k_1)$ plane illustrating the non-monotonic stability behavior of the system, assuming $\tilde{k}_{2}=10^{4}$, $\tilde{\gamma}_{3}=10^{-4}$.}
\end{figure}

Next, we investigate the spectrum for Cosserat rods that can sustain
greater compression, which corresponds to taking smaller values of
$\tilde{k}_{1}$. As illustrated in Fig. \ref{fig:Hopf3}, the parameters
$(\tilde{k}_{1},\tilde{k}_{2},\tilde{\gamma}_{3})=(2\times10^{2},10^{4},10^{-4})$
lead to a qualitatively different dependence of the spectrum on the
follower force. Indeed, while we still observe an onset of damped
oscillations (for $\tilde{\mathcal{F}}\approx22.48$) followed by a Hopf
bifurcation (for $\tilde{\mathcal{F}}\approx49.76$), the window of flutter
instability is now finite, terminating for $\tilde{\mathcal{F}}>150$.
In this regime, which corresponds to a highly compressed Cosserat
rod, we find that the follower force has a destabilizing effect followed
by a stabilizing effect. On the other hand, as illustrated in Fig.
\ref{fig:Hopf4}, the parameters $(\tilde{k}_{1},\tilde{k}_{2},\tilde{\gamma}_{3})=(10^{2},10^{4},10^{-4})$
show only an onset of damped oscillations, and an absence of the Hopf
bifurcation. The bifurcation structure is most sensitive to the variation
of $\tilde{k}_{1}$ and the limit cycle oscillations can be entirely
eliminated by an appropriate choice of this parameter. We illustrate this phenomenon for more values of $\tilde{k}_{1}$ in the phase diagram in  Fig. \ref{fig:phase_schematic}. 
When stretch is permitted, the base state is compressed by the follower force, reducing $\nu = 1 - \tilde{\mathcal{F}}/\tilde{k}_1$. In Section \ref{sec:euler}, we show that the bending term scales as $k_3/\nu^2$, so compression effectively increases the bending stiffness. At sufficiently large $\tilde{\mathcal{F}}$, this compression-induced stiffening dominates the destabilizing circulatory term, causing the eigenvalues to re-enter the stable half-plane and explaining the re-stabilization observed in Fig.  \ref{fig:Hopf3} and Fig. \ref{fig:phase_schematic}.

While this
is a mathematically sound result, our constitutive law assumes linear elasticity even under compression ratios $\nu$ that may become small for large follower forces. In real soft materials, nonlinear strain-stiffening or material buckling would intervene before the $\nu \to 0$ limit is attained. Nevertheless, the stabilizing mechanism identified here relies only on the increase of the effective bend modulus under compression, so we expect the qualitative re-stabilization to persist for more realistic nonlinear constitutive laws, although precise critical loads will shift. A systematic exploration of nonlinear constitutive models is left for future work.

\section{Numerical Simulations \label{sec:simulations}}

Having established the appearance of a linear instability through
a Hopf bifurcation in the previous section, we now carry out fully
nonlinear simulations of the equations of motion to study the dynamics
of Cosserat rods for large follower forces in the unstable regime.
This allows us to determine the fate of the linear instability in
regions where the linear stability analysis is no longer valid. We
employ the compatibility conditions for the simulations as outlined
in Section III. In particular, we solve the nonlinear system of PDEs
in Eq. (\ref{eq:igebmodel}) as a nonlinear time evolution
system for $(\varphi,E)$
\begin{equation}
\partial_{t}\varphi=\varphi V(E),\quad\partial_{t}E_{\alpha}=\mathcal{D}_{\alpha\beta}V_{\beta}(E)
\end{equation}
subject to the boundary conditions in Eq. (\ref{eq:igebBCs}), where
the velocity is eliminated via Eq. (\ref{eq:dynamics}). We non-dimensionalize
the nonlinear system of PDEs by means of the short time scale $\tau_{1}$
and the rest length $L$. We numerically solve the overdamped equations
of motion using a Chebyshev spectral collocation on Mathematica, where
we semi-discretize the partial differential equations in space to
obtain a system of ordinary differential equations in time. These
are solved at the interior nodes simultaneously with the algebraic
equations that result from imposing the boundary conditions at the
two boundary nodes \cite{implementation_note}. The numerical method is validated against the
linear stability analysis and excellent agreement is obtained in the
stable regime of the spectrum for the ranges of constitutive parameters
considered in Section VI. 

In the unstable regime of the spectrum, past the Hopf bifurcation,
we find that the Cosserat rod carries out a self-sustaining beating
motion of constant amplitude corresponding to a limit cycle. We excite
the nonlinear oscillations in the rod by using the first eigenmodes
obtained in the linear stability analysis to form appropriate initial
conditions. For the initial-boundary value problem to be well-posed,
one requires that the initial condition satisfies the nonlinear boundary
conditions of the problem. We circumvent this by introducing a small
dimensionless parameter $\epsilon\ll1$ and imposing the initial data
on $(\varphi,E)$

\begin{align}
\varphi & =\begin{bmatrix}1 & 0 & 0\\
\nu u & \cos\left(\epsilon\,\mathrm{Re}\,\Theta_{1}\right) & -\sin\left(\epsilon\,\mathrm{Re}\,\Theta_{1}\right)\\
\epsilon\,\mathrm{Re}\,Y_{1} & \sin\left(\epsilon\,\mathrm{Re}\,\Theta_{1}\right) & \cos\left(\epsilon\,\mathrm{Re}\,\Theta_{1}\right)
\end{bmatrix},\nonumber \\
E & =\varphi^{-1}\partial_{u}\varphi.
\end{align}
The boundary conditions are then satisfied by the initial data to
first order in the dimensionless parameter $\epsilon$. The limit
cycles are illustrated in Fig. \ref{fig:limitcycle50} for $\tilde{\mathcal{F}}$
for a Cosserat rod with minimal compression that can sustain shear.
We plot the amplitudes of the coordinates of the tip of the rod to
illustrate the tip displacement, as well as snapshots of the entire
Cosserat rod through the simulation. Additionally, for completeness,
we carry out a sweep of the follower force $\tilde{\mathcal{F}}$
to obtain the displacement amplitude of the coordinates of the tip
as a function of $\tilde{\mathcal{F}}$. The results are illustrated
in Fig. \ref{fig:saturation}. 

\onecolumngrid
\begin{widetext}

\begin{figure}[H]
\includegraphics[scale=0.5]{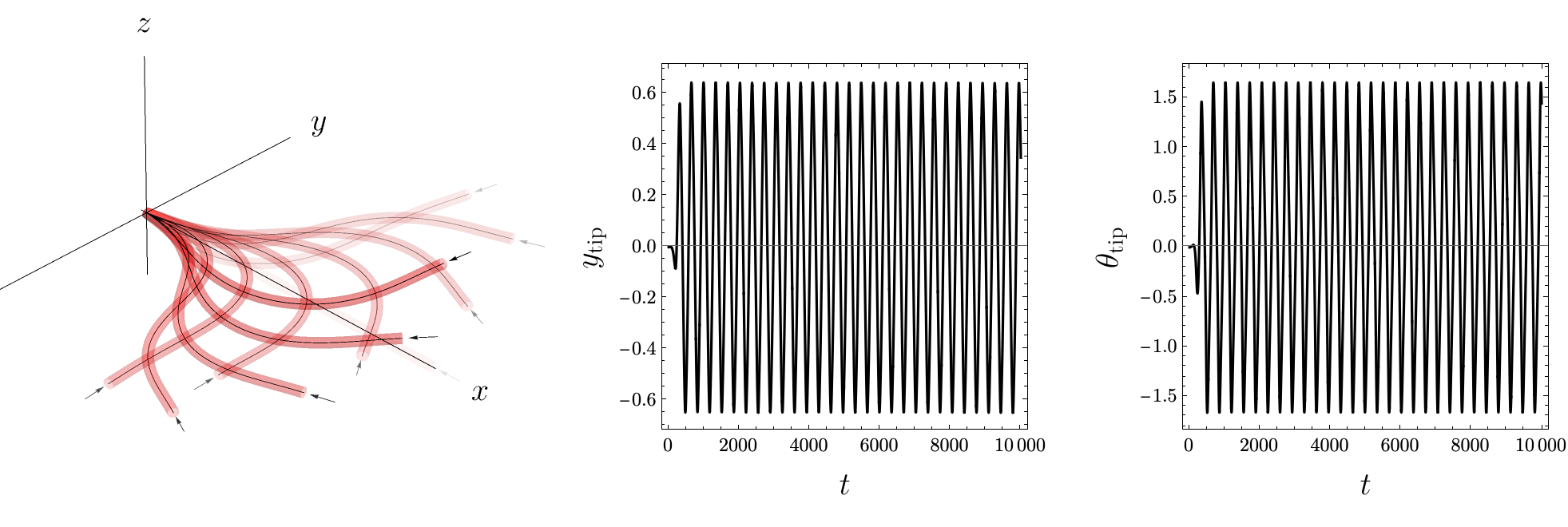}\caption{\label{fig:limitcycle50}Emergence of limit cycle oscillations of
the entire rod (left) and of the top of the rod (right) for $(\tilde{k}_{1},\tilde{k}_{2},\tilde{\gamma}_{3})=(10^{4},10^{2},10^{-2})$
and $\tilde{\mathcal{F}}=50$. Time is measured in units of $\tau_{1}$
in the plots of the tip coordinates. For the visualization of the
rod, the color scale is such that later snapshots of the rod are darker.
The arrows indicate the configuration-dependent direction of the follower
force at each time slice.}
\end{figure}
\begin{figure}[H]
\includegraphics[scale=0.5]{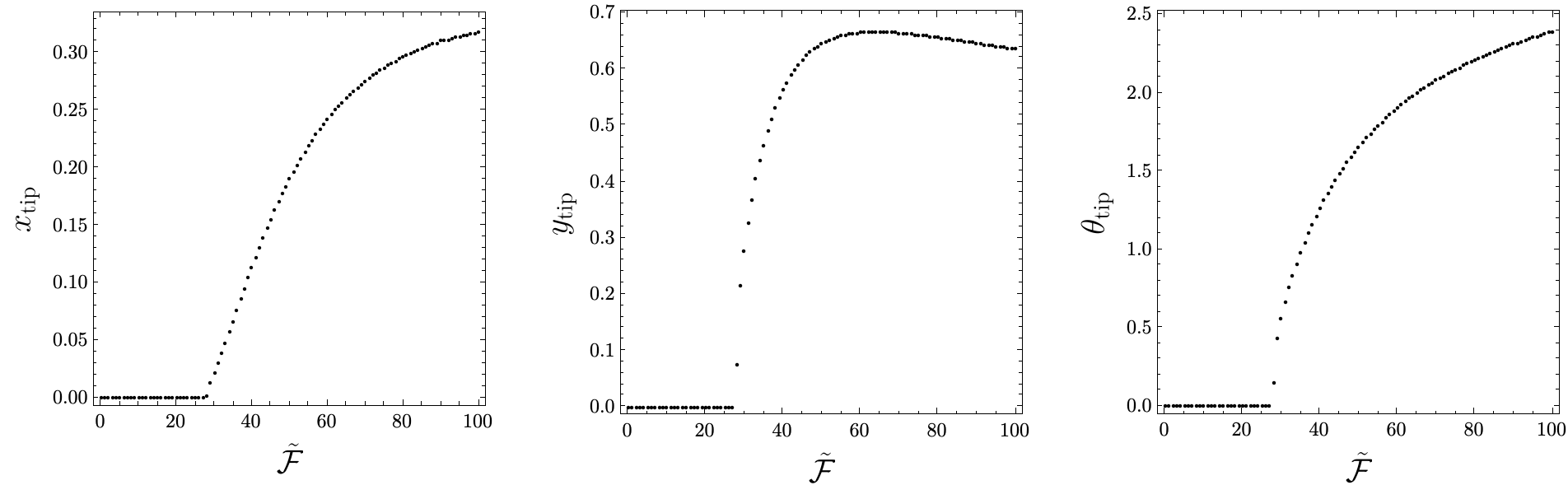}\caption{\label{fig:saturation}The steady-state limit cycle amplitudes of
the coordinates of the tip of the rod as a function of the strength
of the follower force.}
\end{figure}

\end{widetext}
\twocolumngrid
\section{Euler-Bernoulli Limit\label{sec:euler}}

We derive the linearized equations of motion when the time scales
are separated and show that it recovers the Euler-Bernoulli beam limit
considered in the literature. We consider the limit of rapid relaxation
of $\theta^{(1)}$ in which
\begin{equation}
0=k_{3}\partial_{u}^{2}\theta^{(1)}+(k_{2}\nu+\mathcal{F})(\partial_{u}y^{(1)}-\nu\theta^{(1)}).
\end{equation}
Differentiating once gives
\begin{equation}
0=k_{3}\partial_{u}^{3}\theta^{(1)}+(k_{2}\nu+\mathcal{F})\partial_{u}(\partial_{u}y^{(1)}-\nu\theta^{(1)}).
\end{equation}
Thus the equation for $y^{(1)}$ may be written as
\begin{equation}
\gamma_{2}\partial_{t}y^{(1)}=-\frac{k_{2}k_{3}}{(k_{2}\nu+\mathcal{F})}\partial_{u}^{3}\theta^{(1)}-\mathcal{F}\partial_{u}\theta^{(1)}.
\end{equation}
We may then obtain the limit of vanishing shear through taking the
limit $k_{2}\to\infty$ and noting that in this limit,
\begin{equation}
\partial_{u}y^{(1)}=\nu\theta^{(1)}.
\end{equation}
Thus we obtain the closed system of equations for the centerline
\begin{equation}
\begin{aligned}\gamma_{1}\partial_{t}x^{(1)} & =k_{1}\partial_{u}^{2}x^{(1)},\\
\gamma_{2}\partial_{t}y^{(1)} & =-\frac{k_{3}}{\nu^{2}}\partial_{u}^{4}y^{(1)}-\frac{\mathcal{F}}{\nu}\partial_{u}^{2}y^{(1)},
\end{aligned}
\end{equation}
and we note that the $\nu$ scaling increases the effective bending
stiffness of the rod for large values of the follower force. This
intuitively explains why the spectrum of the rod is stabilized when
the stretch degree of freedom is allowed in the system. On the other
hand, in the limit of vanishing stretch $k_{1}\to\infty$, we find
that $\nu=1$ and
\begin{equation}
\gamma_{2}\partial_{t}y^{(1)}=-k_{3}\partial_{u}^{4}y^{(1)}-\mathcal{F}\partial_{u}^{2}y^{(1)},
\end{equation}
which is in agreement with the literature \citep{DeCanio2017,Clarke2024,Schnitzer2025}. In
both of these limits, the elimination of $\theta^{(1)}$ from the
boundary conditions in Eq. (\ref{eq:linearBCs}) leads to the boundary
conditions
\begin{equation}
\begin{aligned}y^{(1)}\lvert_{u=0}=\partial_{u}y^{(1)}\lvert_{u=0} & =0,\\
\partial_{u}^{2}y^{(1)}\lvert_{u=L}=\partial_{u}^{3}y^{(1)}\lvert_{u=L} & =0.
\end{aligned}
\end{equation}

\section{Discussion and Conclusion}

Motivated by recent studies on follower forces in filaments on the
one hand and the geometric field theory literature on the other hand,
we have formulated a geometrically exact model of pressure-driven
soft robotic arms in viscous fluids at low Reynold number. We formulate
our model of a soft robotic arm as a Lie group-valued dissipative
field theory, which leads to compact, coordinate-invariant expressions
for the equations of motion from which convenient coordinatizations
may be derived. In this process, we generalize the classic follower
force model for filaments studied in the biophysics and soft matter
physics literature to Cosserat rods that can sustain shear and stretch. 

We have found that the linearization of this dissipative field theory
leads to non-Hermitian physics in the presence of follower forces
and to oscillatory dynamics. Our choice of constitutive law for the
robotic arm allows us to exploit planar Cosserat degrees of freedom
in their full generality and thus allows us to investigate the impact
of shear, stretch, and bend on the spectrum of the linearized theory.
Our investigations show excellent agreement with the established literature
when these additional degrees of freedom are eliminated - namely,
an onset of oscillatory relaxation followed by a Hopf bifurcation
to stable limit cycle oscillations. On the other hand, departure from
the filament regime leads to spectra that can be substantially different.
We have found that incorporating stretch and shear shifts the critical
values of the follower force monotonically downwards. More crucially,
we have found that the stretch degree of freedom significantly alters
the dynamical state diagram of the model. In the presence of stretch,
the limit cycle is stable only when the magnitude of the follower
force is in a finite interval bounded by a lower and an upper critical
value. Outside this window, no stable oscillations exist and the rod
relaxes to the linear configuration through damped oscillations. We
have also identified a regime in parameter space where the Hopf bifurcation
is absent and only decaying oscillations are obtained. These results
illustrate the subtle nature of incorporating Cosserat degrees of
freedom into the elastohydrodynamics of soft robotic arms. A beam
limit that allows for stretch rationalizes seemingly counterintuitive
phenomenon. 

We carry out geometrically exact, nonlinear simulations of the robotic
arm by leveraging the geometric field theory formalism that we used
to model the problem. We formulate the governing nonlinear PDEs as
a pair of time evolution systems of PDEs that together constitutive
a nonlinear parabolic-hyperbolic system of equations. We leverage
the Lie-algebraic structure of velocities, deformations, and the covariant
representation of the constitutive law to write down a time evolution
equation for the deformation gradient $E$ entirely in the Lie algebra.
This is decoupled from another time evolution equation for our Cosserat
field $\varphi$ in the Lie group, which allows us to reconstruct and
visualize configurations of the robotic arm as necessary. Our geometrically
exact numerical integration scheme illustrates that in the linearly
unstable regime of the spectrum, one obtains limit cycle oscillations
of the robotic arm where the arm exhibits constant amplitude self-sustaining
oscillations. There is excellent agreement between the analytical
theory and the numerical results. 

We highlight some future extensions of our present work. In a future
publication, we shall carry out a perturbative expansion of the dissipative
field theory to study the nature of the emergence of the limit cycle
near the Hopf bifurcation. This allows us to analytically predict
the amplitude of the limit cycle near the Hopf bifurcation. On the
other hand, we shall also extend our geometric field theory formalism
to three spatial dimensions for soft robotic arms that can also sustain
torsion and twist. One may also investigate the effect of follower
torques in three-dimensional models of soft robotic arms. Finally,
one may generalize this dissipative field theory to arbitrary Lie
groups $G$ to study the dissipative mechanics of complex systems
with Lie group-valued degrees of freedom. This would complement the
conservative mechanics of such systems which have been studied extensively
in the literature as $G$-snakes \citep{Krishnaprasad1994} and $G$-strands
\citep{Holm2012}. 
\begin{acknowledgments}
We thank Professors M. E. Cates, R. E. Goldstein, and D. D. Holm for
insightful discussions concerning mechanics and geometry. We thank
Dr. Lukas Kikuchi, Bal\'{a}zs N\'{e}meth, Mingjia Yan, and Hanchun Wang and
for many helpful suggestions. We thank Professor M. C. Payne and Dr.
A. Souslov for facilitating a collaboration between the Cavendish
Laboratory and the Department of Applied Mathematics and Theoretical
Physics. Mohamed Warda acknowledges a PhD studentship from the EPSRC (award number W108115D) and UKRI.
\end{acknowledgments}

\appendix

\section{Geometric Conventions\label{sec:geometry}}

We review the important properties associated with the geometric manipulations
carried out in this paper. More details may be found in \citep{Yan2025}.
We use Greek indices throughout the paper for components of quantities
that are in $\mathfrak{se}(2)$ or its dual, $\mathfrak{se}(2)^{\star}$.
We represent elements $X\in\mathfrak{se}(2)$ as the $3\times3$ matrices
\begin{equation}
X=\begin{bmatrix}0 & 0 & 0\\
\zeta_{1} & 0 & -\Lambda\\
\zeta_{2} & \Lambda & 0
\end{bmatrix}=X_{\alpha}b_{\alpha},
\end{equation}
where $X_{\alpha}=(\zeta_{1},\zeta_{2},\Lambda)$ and $b_{\alpha}$
are the generators of the Lie algebra, which form a canonical basis
of $\mathfrak{se}(2)$ and an associated dual basis $b_{\alpha}^{\star}$
for $\mathfrak{se}(2)^{*}$. The algebra of $\mathfrak{se}(2)$ is equipped with
a Lie bracket $[\cdot,\cdot]$ which in our case is a matrix commutator
$[X,Y]=XY-YX$. The structure constants $C_{\alpha\beta\mu}$ of $\mathfrak{se}(2)$
relative to its canonical basis are determined via the relations $[b_{\alpha},b_{\beta}]=C_{\alpha\beta\mu}b_{\mu}$,
which are given explicitly by
\begin{equation}
[b_{1},b_{2}]=0,\quad[b_{3},b_{1}]=b_{2},\quad[b_{3},b_{2}]=-b_{1}.
\end{equation}
The adjoint map is defined as 
\begin{equation}
\mathrm{ad}_{X}Y=[X,Y].
\end{equation}
We have a natural pairing between $\mathfrak{se}(2)^{*}$ and $\mathfrak{se}(2)$,
$\langle\cdot,\cdot\rangle$, defined for $\sigma\in\mathfrak{se}(2)^{*}$,
$X\in\mathfrak{se}(2)$ by
\begin{equation}
\langle\sigma,X\rangle=\sigma_{\alpha}X_{\alpha},\label{eq:pairing}
\end{equation}
through which we define the coadjoint $\mathrm{ad}^{*}$ 
\begin{equation}
\langle\mathrm{ad}_{X}^{*}\sigma,Y\rangle=-\langle\sigma,\mathrm{ad}_{X}Y\rangle.
\end{equation}
We note that for concrete calculations, the component-wise form of
the pairing $\langle\cdot,\cdot\rangle$ is more useful. On the other
hand, it is more convenient to preserve the component-free form of
the pairing for more theoretical manipulations that are occasionally
carried out in the paper. 

We define the covariant derivatives
\begin{equation}
\begin{aligned}\mathcal{D} & =\partial_{u}+\mathrm{ad}_{E},\\
\mathcal{D}^{*} & =\partial_{u}+\mathrm{ad}_{E}^{*},
\end{aligned}
\end{equation}
which, by construction, satisfy the Leibniz rule
\begin{equation}
\partial_{u}\langle\sigma,X\rangle=\langle\mathcal{D}^{*}\sigma,X\rangle+\langle\sigma,\mathcal{D}X\rangle.
\end{equation}
For concrete computations we make use of the canonical basis isomorphism
in which we work relative to the basis $b_{\alpha}$ of $\mathfrak{se}(2)$,
implicitly taking elements of both the Lie algebra and its dual to
be column vectors in $\mathbb{R}^{3}$ (or $\mathbb{C}^{3}$ when
studying the spectra of operators). For such computations, we find
the matrix representations of the $\mathrm{ad}$ and $\mathrm{ad}^{*}$
operations to be
\begin{equation}
(\mathrm{ad}_{X})_{\alpha\beta}=\begin{bmatrix}0 & -\Lambda & \zeta_{2}\\
\Lambda & 0 & -\zeta_{1}\\
0 & 0 & 0
\end{bmatrix},\quad(\mathrm{ad}_{X}^{*})_{\alpha\beta}=-(\mathrm{ad}_{X})_{\beta\alpha},
\end{equation}
and we work with the covariant derivatives as differential operator-valued
matrices
\begin{equation}
\mathcal{D}_{\alpha\beta}=\delta_{\alpha\beta}\partial_{u}+(\mathrm{ad}_{E})_{\alpha\beta},\quad\mathcal{D}_{\alpha\beta}^{*}=\delta_{\alpha\beta}\partial_{u}+(\mathrm{ad}_{E}^{*})_{\alpha\beta}.
\end{equation}
For our particular set up of overdamped problems we are provided with
the friction tensor $\Gamma:\mathfrak{se}(2)\to\mathfrak{se}(2)^{*}$ which
we may interpret as a metric that allows us to induce an inner product
on $\mathfrak{se}(2)$ itself via
\begin{equation}
X\cdot Y=\langle\Gamma X,Y\rangle=\Gamma_{\alpha\beta}X_{\beta}Y_{\alpha}.
\end{equation}
This additionally allows us to define an inner product on the Hilbert
space of functions valued in the Lie algebra, which is discussed in
Appendix \ref{sec:hermiticity}. 

\section{Properties of $\mathcal{L}$\label{sec:hermiticity}}

We work with the $\Gamma-$inner product defined on a Hilbert space
of functions $\xi_{\alpha}:[0,L]\to\mathbb{C}^{3}$
\begin{equation}
(\psi,\xi)_{\Gamma}=\int_{0}^{L}\mathrm{d}u\,\,\langle\Gamma\psi,\xi\rangle,
\end{equation}
where we extend the pairing in Eq. (\ref{eq:pairing}) to allow for
complex coefficients $\langle\Gamma\psi,\xi\rangle=\Gamma_{\alpha\beta}\psi_{\beta}^{*}\xi_{\alpha}$.
It is straight forward to verify $\langle\Gamma\psi,\xi\rangle=\langle\Gamma\xi,\psi\rangle^{*}$
and $\langle K\psi,\xi\rangle=\langle K\xi,\psi\rangle^{*}$, which
we implicitly use in the calculation below. 

We obtain $\mathcal{L}^{\dagger}$ through its definition $(\psi,\mathcal{L}\xi)_{\Gamma}=(\mathcal{L}^{\dagger}\psi,\xi)_{\Gamma}$
and its convenient expression in the form
\begin{equation}
\mathcal{L}\xi=\Gamma^{-1}(\mathcal{D}^{*}(K\mathcal{D}\xi)+\mathrm{ad}_{\mathcal{D}\xi}^{*}\Sigma^{(0)}),
\end{equation}
where
\begin{equation}
\mathcal{D}=\partial_u+\mathrm{ad}_{E^{(0)}}.
\end{equation}
We find through the properties of $\mathrm{ad}$ and $\mathcal{D}$
\begin{align}
(\psi,\mathcal{L}\xi)_{\Gamma} & =\int_{0}^{L}\mathrm{d}u\,\,\left\langle \mathcal{D}^{*}(K\mathcal{D}\xi)+\mathrm{ad}_{\mathcal{D}\xi}^{\star}\Sigma^{(0)},\psi\right\rangle ^{*}\nonumber \\
 & =\left\langle K\mathcal{D}\xi,\psi\right\rangle ^{*}\bigg\lvert_{u=0}^{u=L}-\left\langle K\mathcal{D}\psi+\mathrm{ad}_{\psi}^{*}\Sigma^{(0)},\xi\right\rangle \bigg\lvert_{u=0}^{u=L}\nonumber \\
 & +\int_{0}^{L}\mathrm{d}u\,\,\left\langle \left(\mathcal{D}^{*}(K\mathcal{D}\psi)+\mathrm{ad}_{\mathcal{D}\psi}^{*}\Sigma^{(0)}\right),\xi\right\rangle \nonumber \\
 & =\left\langle K\mathcal{D}\xi,\psi\right\rangle ^{*}\bigg\lvert_{u=0}^{u=L}-\left\langle K\mathcal{D}\psi+\mathrm{ad}_{\psi}^{*}\Sigma^{(0)},\xi\right\rangle \bigg\lvert_{u=0}^{u=L}\nonumber \\
 & +(\mathcal{L}^{\dagger}\psi,\xi)_{\Gamma},
\end{align}
where $\mathcal{L}^{\dagger}$ is operationally identical to$\mathcal{L}$
\begin{equation}
\mathcal{L}^{\dagger}\psi=\Gamma^{-1}(\mathcal{D}^{*}(K\mathcal{D}\psi)+\mathrm{ad}_{\mathcal{D}\psi}^{*}\Sigma^{(0)}).
\end{equation}
Imposing that the boundary terms vanish and using the
primal boundary conditions
\begin{equation}
\xi\lvert_{u=0}=0,\quad\mathcal{D}\xi\lvert_{u=L}=0
\end{equation}
produces the adjoint boundary conditions
\begin{equation}
\psi\lvert_{u=0}=0,\quad\left(K\mathcal{D}\psi+\mathrm{ad}_{\psi}^{*}\Sigma^{(0)}\right)\big\lvert_{u=L}=0,
\end{equation}
which are different from the primal boundary conditions for $\mathcal{F}\neq0$.
Thus we see that for $\mathcal{F}\neq0$, $\mathcal{L}\neq\mathcal{L}^{\dagger}$
since they act on different domains, and the system exhibits a loss
of Hermiticity. On the other hand, for $\mathcal{F}=0$, the domains
of the two operators $\mathcal{L}$ and $\mathcal{L}^{\dagger}$ are
identical, and we obtain $\mathcal{L}=\mathcal{L}^{\dagger}$ - in
this case, $\mathcal{L}$ is Hermitian and admits strictly real eigenvalues.
Moreover, in the case that $\mathcal{F}=0$, we immediately see that
$\mathcal{L}$ is necessarily a negative definite operator. Indeed,
this is another manifestation of the conservation of energy
\begin{equation}
\begin{aligned}(\xi,\mathcal{L}\xi)_{\Gamma} & =\int_{0}^{L}\mathrm{d}u\,\,\left\langle \xi,\mathcal{D}^{*}(K\mathcal{D}\xi)\right\rangle \\
 & =-\int_{0}^{L}\mathrm{d}u\,\,\left\langle K\mathcal{D}\xi,\mathcal{D}\xi\right\rangle \\
 & \leq0.
\end{aligned}
\end{equation}

\section{Eigenvalues and Eigenvectors\label{sec:eigenvalues}}

We introduce the auxiliary quantity $\mu=\tilde{k}_{2}\nu+\tilde{\mathcal{F}}$
for convenience. The characteristic polynomial for the transverse
subsystem corresponding to solutions of the form
\begin{equation}
\Xi_{\alpha}\sim\begin{bmatrix}0\\
A_{1}\\
A_{2}
\end{bmatrix}\exp(\omega t)\exp(qu)
\end{equation}
may be written, after non-dimensionalization, as the quartic equation
in $q$
\begin{equation}
\frac{\tilde{k}_{2}}{\tilde{\gamma}_{3}}q^{4}+\left(\tilde{\mathcal{F}}\mu-\left(\tilde{k}_{2}+\frac{1}{\tilde{\gamma}_{3}}\right)\omega\right)q^{2}+\omega(\omega+\nu\mu)=0.
\end{equation}
Thus we find that for a given value of $\omega$, there are four associated
roots $q\in\{\pm\chi_{1},\pm i\chi_{2}\}$ of the quartic. Note that
this convention is inspired by the case of $\tilde{\mathcal{F}}=0$,
where $\omega<0$ and $\chi_{1},\chi_{2}\in\mathbb{R}$. Thus, we
may write
\begin{equation}
\begin{aligned}\Xi(u) & =C_{1}\begin{bmatrix}0\\
\cosh\chi_{1}u\\
g_{1}\sinh\chi_{1}u
\end{bmatrix}+C_{2}\begin{bmatrix}0\\
\sinh\chi_{1}u\\
g_{1}\cosh\chi_{1}u
\end{bmatrix}\\
 & +C_{3}\begin{bmatrix}0\\
\cos\chi_{2}u\\
-g_{2}\sin\chi_{2}u
\end{bmatrix}+C_{4}\begin{bmatrix}0\\
-\sin\chi_{2}u\\
-g_{2}\cos\chi_{2}u
\end{bmatrix},
\end{aligned}
\end{equation}
where $C_{1},C_{2},C_{3},C_{4}\in\mathbb{C}$ are constants and $g_{1},g_{2}$
are the auxiliary quantities
\begin{equation}
g_{1}=\frac{\tilde{k}_{2}\chi_{1}^{2}-\omega}{\mu\chi_{1}},\quad g_{2}=\frac{\tilde{k}_{2}\chi_{2}^{2}+\omega}{\mu\chi_{2}}.
\end{equation}
The boundary conditions
\begin{equation}
\begin{aligned} & Y(0)=\Theta(0)=0,\\
 & Y'(1)-\nu\Theta(1)=\Theta'(1)=0,
\end{aligned}
\end{equation}
lead to a $4\times4$ system of linear equations for the constants
$C_{i}$, which may be written as $\Delta_{ij}C_{j}=0$ where $\Delta$
is a $4\times4$ matrix. The condition $\mathrm{det}\,\Delta=0$ simplifies
to the following transcendental equation for $\omega$
\begin{equation}
B_{1}+B_{2}\sinh\chi_{1}\sin\chi_{2}+B_{3}\cosh\chi_{1}\cos\chi_{2}=0,
\end{equation}
where 
\begin{equation}
\begin{aligned}B_{1} & =(\chi_{1}(\chi_{1}-\nu g_{1})-\chi_{2}(\chi_{2}-\nu g_{2}))g_{1},\\
B_{2} & =-(\chi_{2}(\chi_{1}-\nu g_{1})+\chi_{1}(\chi_{2}-\nu g_{2}))g_{1},\\
B_{3} & =\frac{\chi_{2}g_{2}^{2}(\chi_{1}-\nu g_{1})-\chi_{1}g_{1}^{2}(\chi_{2}-\nu g_{2})}{g_{2}}.
\end{aligned}
\end{equation}
The transcendental equation admits countably infinitely many solutions $\omega$
in the complex numbers for a given choice of constitutive parameters
and follower force. For each such $\omega$, we compute its associated
eigenmode $\Xi=(0,Y,\Theta)$

\begin{widetext}
\begin{equation}
\begin{aligned}Y(u) & =\mathcal{N}\left(\cosh\chi_{1}u-\cos\chi_{2}u-G\left(\frac{1}{g_{1}}\sinh\chi_{1}u-\frac{1}{g_{2}}\sin\chi_{2}u\right)\right),\\
\Theta(u) & =\mathcal{N}\left(g_{1}\sinh\chi_{1}u+g_{2}\sin\chi_{2}u-G\left(\cosh\chi_{1}u-\cos\chi_{2}u\right)\right),
\end{aligned}
\end{equation}

\end{widetext}where
\begin{equation}
G=\frac{\chi_{1}g_{1}\cosh\chi_{1}+\chi_{2}g_{2}\cos\chi_{2}}{\chi_{1}\sinh\chi_{1}+\chi_{2}\sin\chi_{2}}
\end{equation}
and $\mathcal{N}$ is a choice of normalization. We choose $\mathcal{N}$
such that $Y(1)=1$.

\onecolumngrid

\newpage 

\begin{figure}[H]
\subfloat[Eigenmodes]{\includegraphics[scale=0.4]{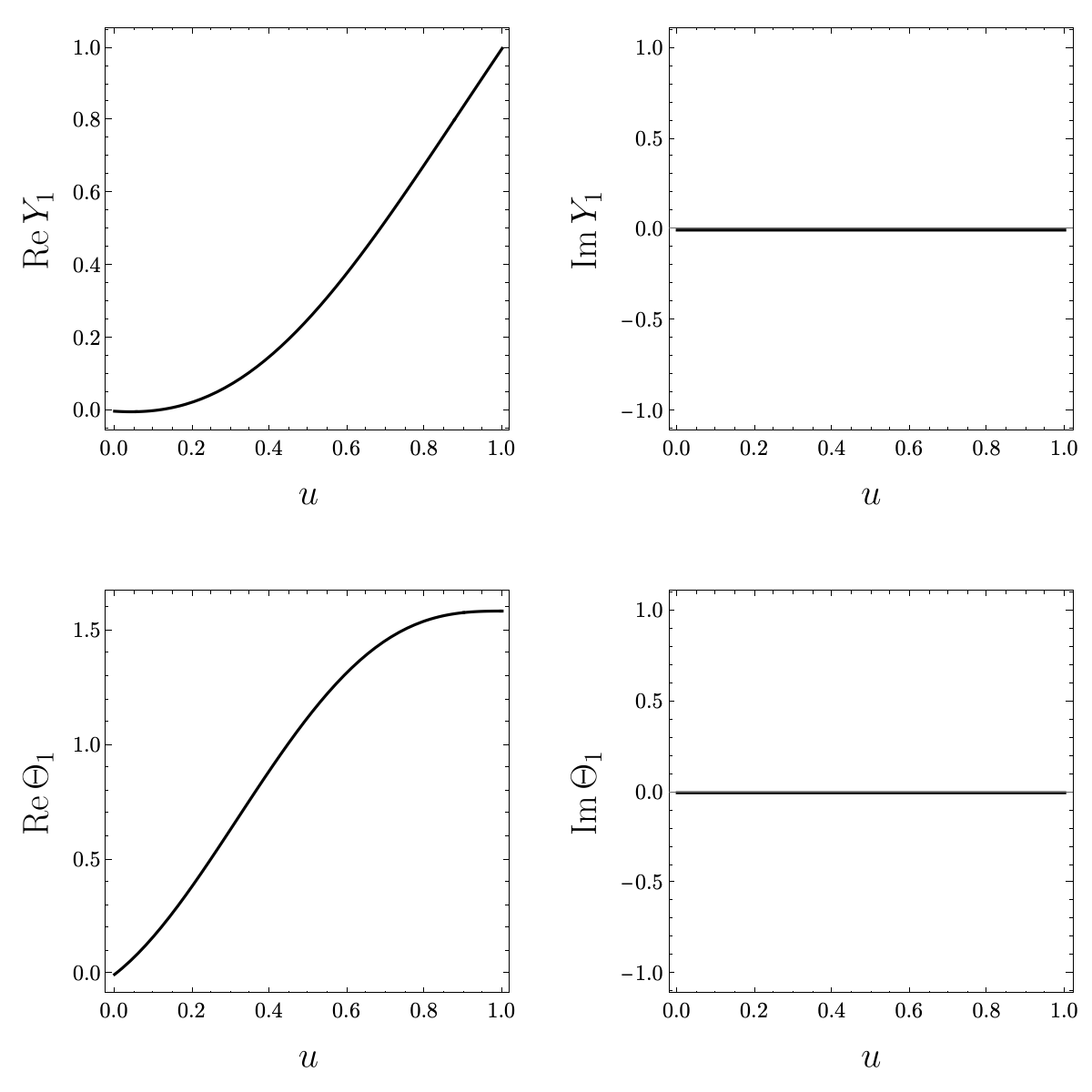}}\subfloat[Deformations]{\includegraphics[scale=0.4]{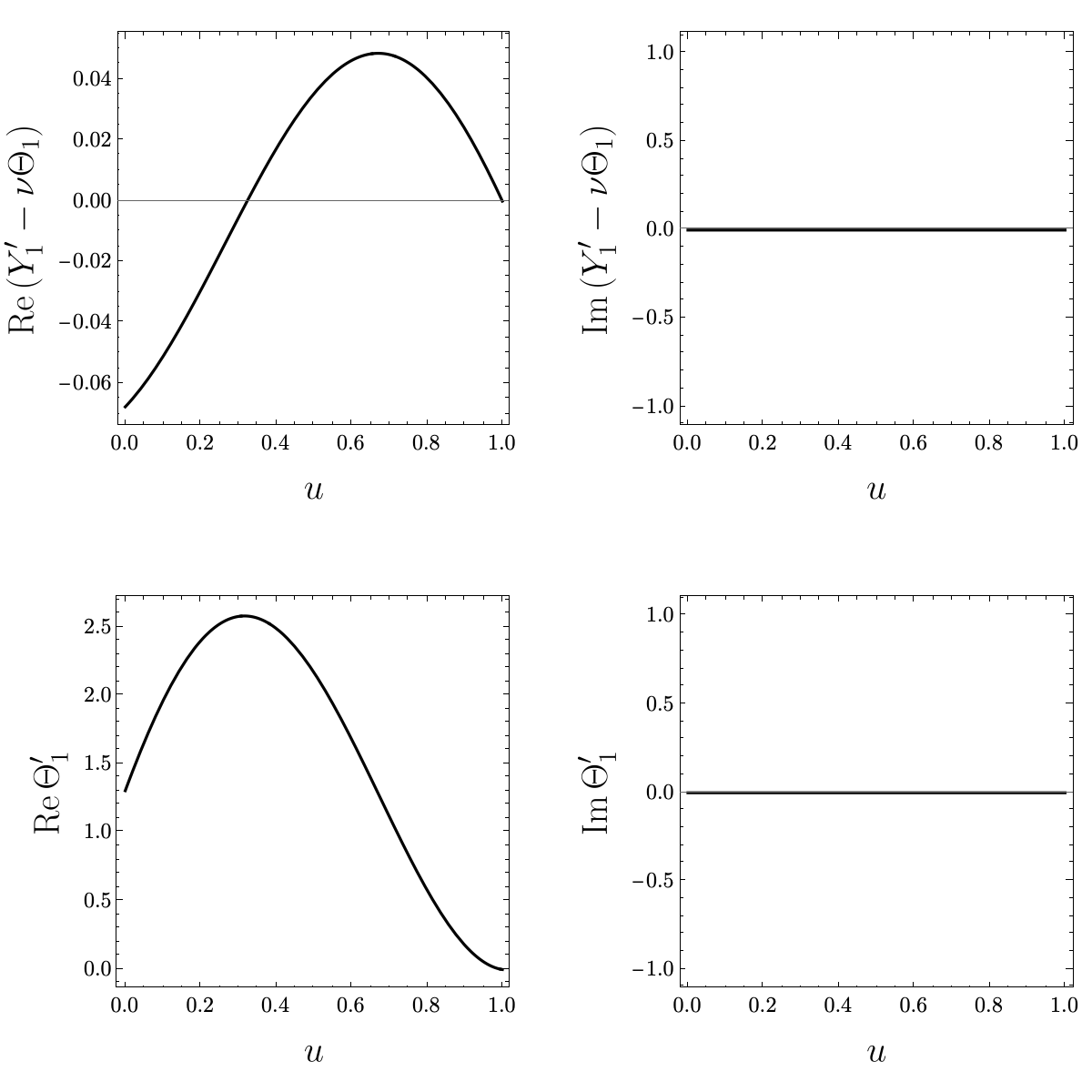}}\caption{\label{fig:eigenmodes10}The first eigenmode $(0,Y_{1},\Theta_{1})$
and its associated deformations for $(\tilde{k}_{1},\tilde{k}_{2},\tilde{\gamma}_{3})=(10^{4},10^{2},10^{-2})$
and $\tilde{\mathcal{F}}=10$, before the onset of oscillations.}
\end{figure}

\begin{figure}[H]
\subfloat[Eigenmodes]{\includegraphics[scale=0.4]{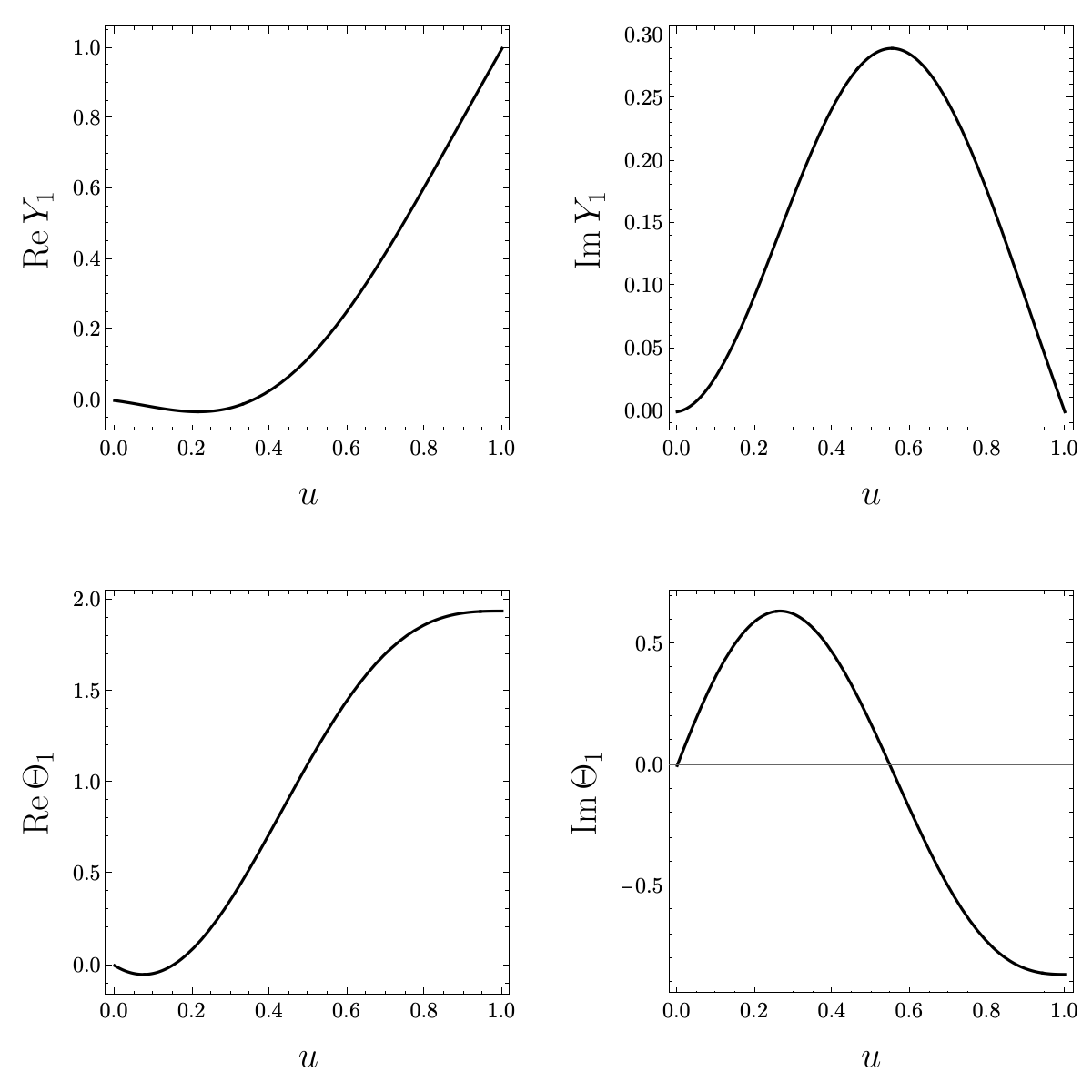}}\subfloat[Deformations]{\includegraphics[scale=0.4]{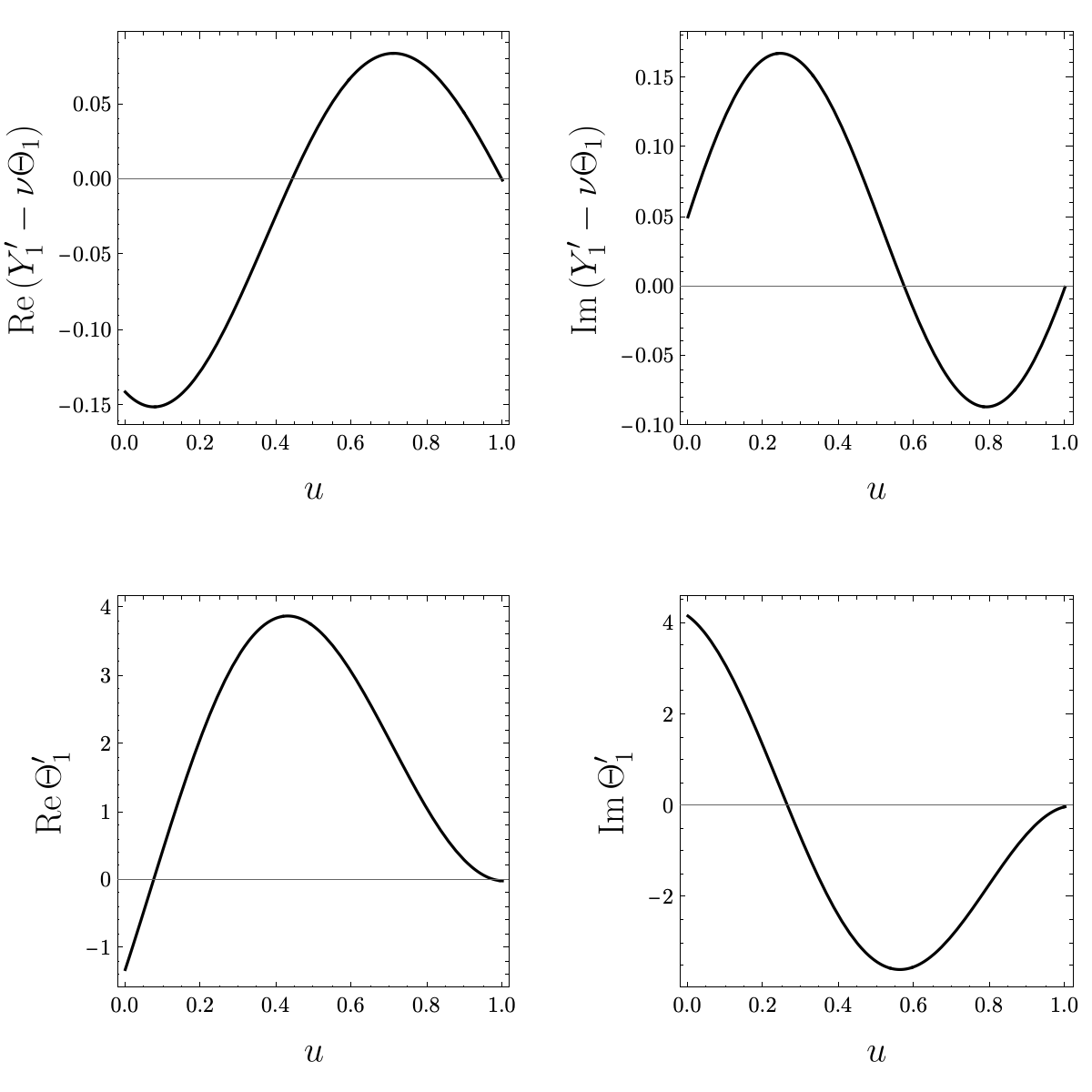}}\caption{\label{fig:eigenmodes20}The first eigenmode $(0,Y_{1},\Theta_{1})$
and its associated deformations for $(\tilde{k}_{1},\tilde{k}_{2},\tilde{\gamma}_{3})=(10^{4},10^{2},10^{-2})$
and $\tilde{\mathcal{F}}=20$, after the onset of oscillations but
before the Hopf bifurcation.}
\end{figure}

\begin{figure}[H]
\subfloat[Eigenmodes]{\includegraphics[scale=0.4]{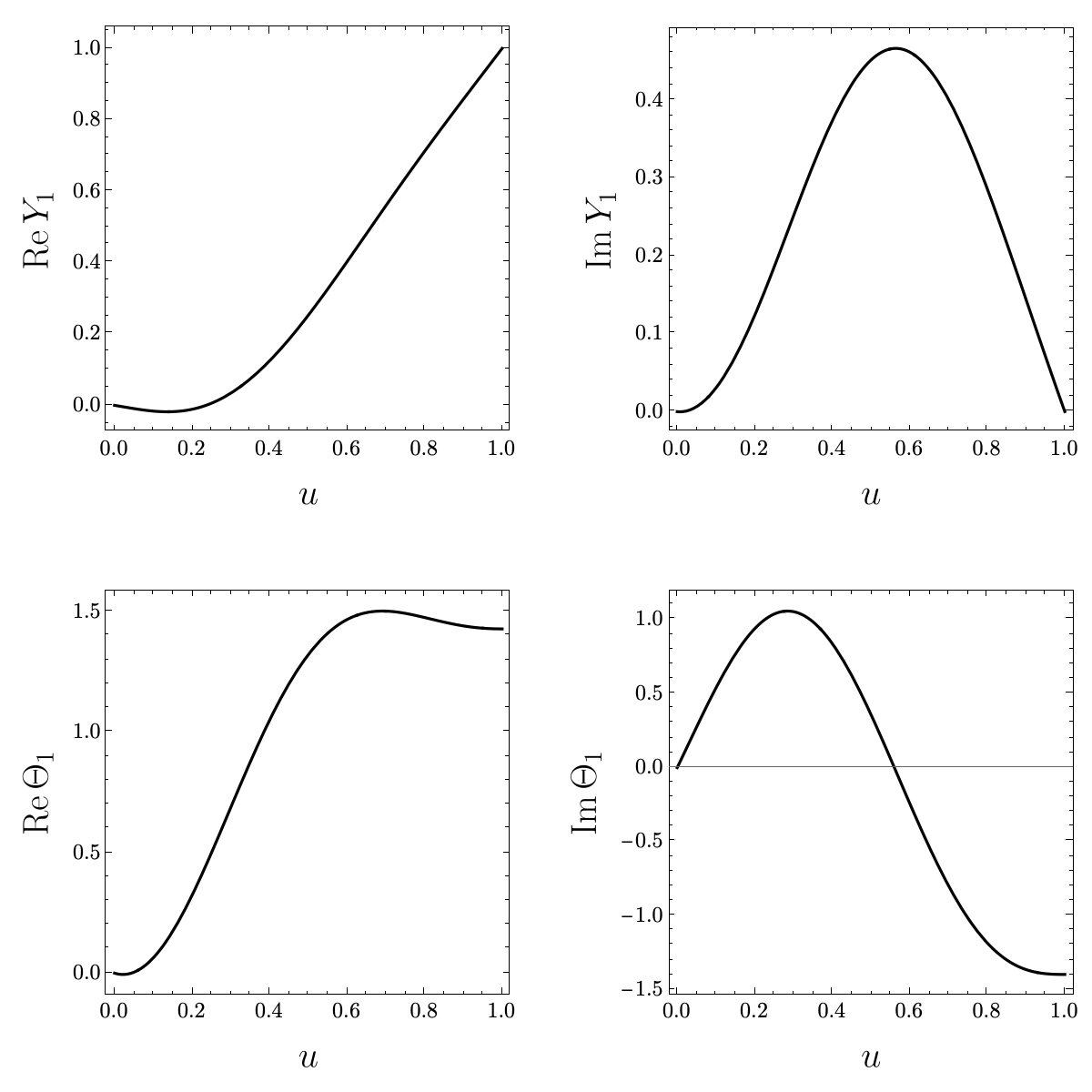}}\subfloat[Deformations]{\includegraphics[scale=0.4]{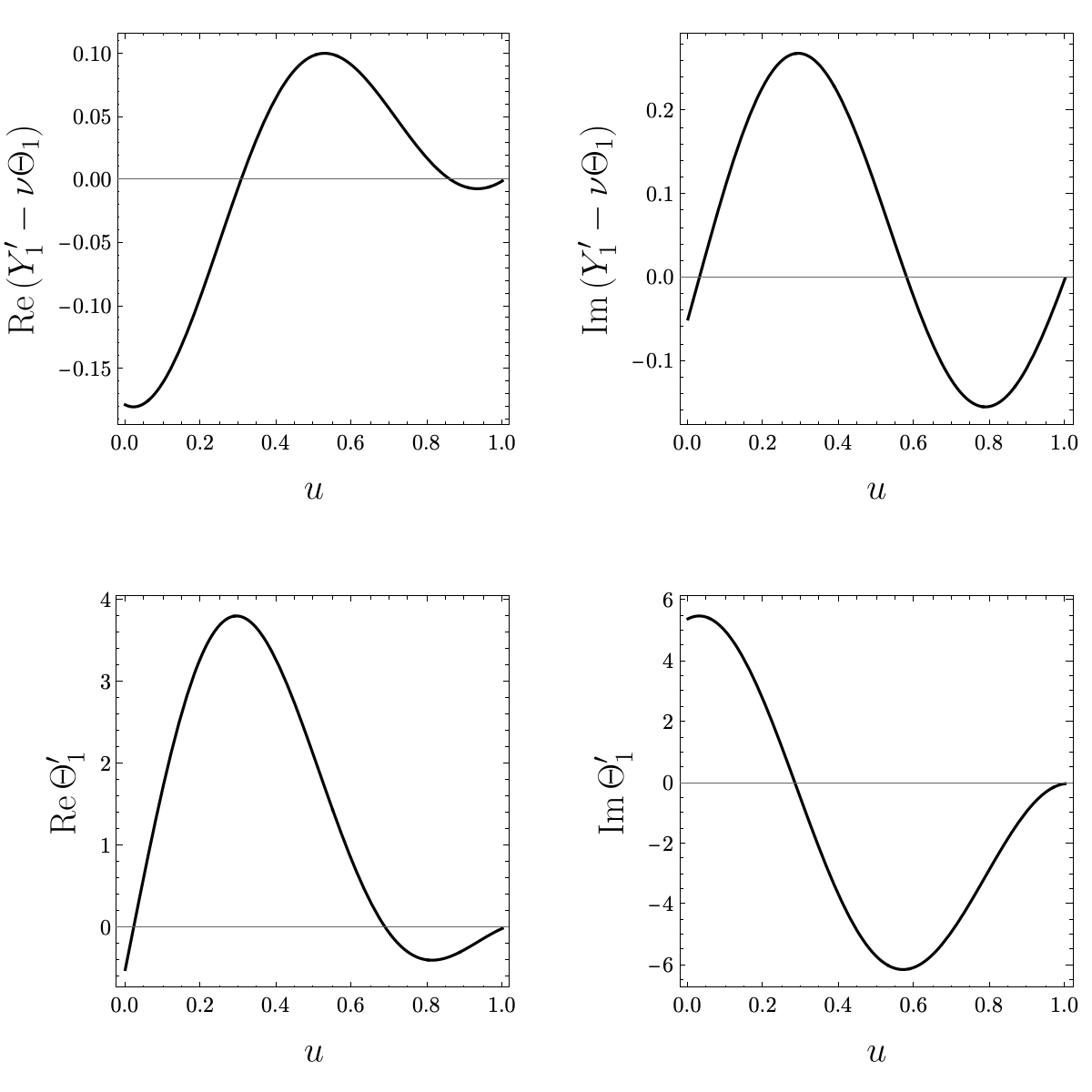}}\caption{\label{fig:eigenmodes30}The first eigenmode $(0,Y_{1},\Theta_{1})$
and its associated deformations for $(\tilde{k}_{1},\tilde{k}_{2},\tilde{\gamma}_{3})=(10^{4},10^{2},10^{-2})$
and $\tilde{\mathcal{F}}=30$, after the Hopf bifurcation.}
\end{figure}


\twocolumngrid

\bibliography{bibliography}

\begin{thebibliography}{39}%
\makeatletter
\providecommand \@ifxundefined [1]{%
 \@ifx{#1\undefined}
}%
\providecommand \@ifnum [1]{%
 \ifnum #1\expandafter \@firstoftwo
 \else \expandafter \@secondoftwo
 \fi
}%
\providecommand \@ifx [1]{%
 \ifx #1\expandafter \@firstoftwo
 \else \expandafter \@secondoftwo
 \fi
}%
\providecommand \natexlab [1]{#1}%
\providecommand \enquote  [1]{``#1''}%
\providecommand \bibnamefont  [1]{#1}%
\providecommand \bibfnamefont [1]{#1}%
\providecommand \citenamefont [1]{#1}%
\providecommand \href@noop [0]{\@secondoftwo}%
\providecommand \href [0]{\begingroup \@sanitize@url \@href}%
\providecommand \@href[1]{\@@startlink{#1}\@@href}%
\providecommand \@@href[1]{\endgroup#1\@@endlink}%
\providecommand \@sanitize@url [0]{\catcode `\\12\catcode `\$12\catcode `\&12\catcode `\#12\catcode `\^12\catcode `\_12\catcode `\%12\relax}%
\providecommand \@@startlink[1]{}%
\providecommand \@@endlink[0]{}%
\providecommand \url  [0]{\begingroup\@sanitize@url \@url }%
\providecommand \@url [1]{\endgroup\@href {#1}{\urlprefix }}%
\providecommand \urlprefix  [0]{URL }%
\providecommand \Eprint [0]{\href }%
\providecommand \doibase [0]{https://doi.org/}%
\providecommand \selectlanguage [0]{\@gobble}%
\providecommand \bibinfo  [0]{\@secondoftwo}%
\providecommand \bibfield  [0]{\@secondoftwo}%
\providecommand \translation [1]{[#1]}%
\providecommand \BibitemOpen [0]{}%
\providecommand \bibitemStop [0]{}%
\providecommand \bibitemNoStop [0]{.\EOS\space}%
\providecommand \EOS [0]{\spacefactor3000\relax}%
\providecommand \BibitemShut  [1]{\csname bibitem#1\endcsname}%
\let\auto@bib@innerbib\@empty
\bibitem [{\citenamefont {Murray}(1994)}]{Murray1994}%
  \BibitemOpen
  \bibfield  {author} {\bibinfo {author} {\bibfnamefont {R.~M.}\ \bibnamefont {Murray}},\ }\href@noop {} {\emph {\bibinfo {title} {A {Mathematical} {Introduction} to {Robotic} {Manipulation}}}}\ (\bibinfo  {publisher} {CRC Press},\ \bibinfo {address} {Boca Raton},\ \bibinfo {year} {1994})\BibitemShut {NoStop}%
\bibitem [{\citenamefont {Selig}(2004)}]{Selig2004}%
  \BibitemOpen
  \bibfield  {author} {\bibinfo {author} {\bibfnamefont {J.~M.}\ \bibnamefont {Selig}},\ }\href@noop {} {\emph {\bibinfo {title} {Geometric {Fundamentals} of {Robotics}}}}\ (\bibinfo  {publisher} {Springer},\ \bibinfo {address} {New York},\ \bibinfo {year} {2004})\BibitemShut {NoStop}%
\bibitem [{\citenamefont {Chirikjian}(2009)}]{Chirikjian2009}%
  \BibitemOpen
  \bibfield  {author} {\bibinfo {author} {\bibfnamefont {G.~S.}\ \bibnamefont {Chirikjian}},\ }\href {https://doi.org/10.1007/978-0-8176-4803-9} {\emph {\bibinfo {title} {Stochastic Models, Information Theory, and Lie Groups, Volume 1: Classical Results and Geometric Methods}}}\ (\bibinfo  {publisher} {Birkh\"{a}user Boston},\ \bibinfo {year} {2009})\BibitemShut {NoStop}%
\bibitem [{\citenamefont {Chirikjian}\ and\ \citenamefont {Burdick}(1994)}]{Chirikjian1994}%
  \BibitemOpen
  \bibfield  {author} {\bibinfo {author} {\bibfnamefont {G.~S.}\ \bibnamefont {Chirikjian}}\ and\ \bibinfo {author} {\bibfnamefont {J.~W.}\ \bibnamefont {Burdick}},\ }\bibfield  {title} {\bibinfo {title} {A hyper-redundant manipulator},\ }\href@noop {} {\bibfield  {journal} {\bibinfo  {journal} {IEEE Robot. Autom. Mag.}\ }\textbf {\bibinfo {volume} {1}},\ \bibinfo {pages} {22} (\bibinfo {year} {1994})}\BibitemShut {NoStop}%
\bibitem [{\citenamefont {Ranzani}\ \emph {et~al.}(2016)\citenamefont {Ranzani}, \citenamefont {Cianchetti}, \citenamefont {Gerboni}, \citenamefont {Falco},\ and\ \citenamefont {Menciassi}}]{Ranzani2016}%
  \BibitemOpen
  \bibfield  {author} {\bibinfo {author} {\bibfnamefont {T.}~\bibnamefont {Ranzani}}, \bibinfo {author} {\bibfnamefont {M.}~\bibnamefont {Cianchetti}}, \bibinfo {author} {\bibfnamefont {G.}~\bibnamefont {Gerboni}}, \bibinfo {author} {\bibfnamefont {I.~D.}\ \bibnamefont {Falco}},\ and\ \bibinfo {author} {\bibfnamefont {A.}~\bibnamefont {Menciassi}},\ }\bibfield  {title} {\bibinfo {title} {A soft modular manipulator for minimally invasive surgery: Design and characterization of a single module},\ }\href@noop {} {\bibfield  {journal} {\bibinfo  {journal} {IEEE Trans. Robot.}\ }\textbf {\bibinfo {volume} {32}},\ \bibinfo {pages} {187} (\bibinfo {year} {2016})}\BibitemShut {NoStop}%
\bibitem [{\citenamefont {Campisano}\ \emph {et~al.}(2020)\citenamefont {Campisano}, \citenamefont {Remirez}, \citenamefont {Cal\'{o}}, \citenamefont {Chandler}, \citenamefont {Obstein}, \citenamefont {Webster},\ and\ \citenamefont {Valdastri}}]{Campisano2020}%
  \BibitemOpen
  \bibfield  {author} {\bibinfo {author} {\bibfnamefont {F.}~\bibnamefont {Campisano}}, \bibinfo {author} {\bibfnamefont {A.~A.}\ \bibnamefont {Remirez}}, \bibinfo {author} {\bibfnamefont {S.}~\bibnamefont {Cal\'{o}}}, \bibinfo {author} {\bibfnamefont {J.~H.}\ \bibnamefont {Chandler}}, \bibinfo {author} {\bibfnamefont {K.~L.}\ \bibnamefont {Obstein}}, \bibinfo {author} {\bibfnamefont {R.~J.}\ \bibnamefont {Webster}},\ and\ \bibinfo {author} {\bibfnamefont {P.}~\bibnamefont {Valdastri}},\ }\bibfield  {title} {\bibinfo {title} {Online disturbance estimation for improving kinematic accuracy in continuum manipulators},\ }\href@noop {} {\bibfield  {journal} {\bibinfo  {journal} {IEEE Robot. Autom. Lett.}\ }\textbf {\bibinfo {volume} {5}},\ \bibinfo {pages} {2642} (\bibinfo {year} {2020})}\BibitemShut {NoStop}%
\bibitem [{\citenamefont {Campisano}\ \emph {et~al.}(2021)\citenamefont {Campisano}, \citenamefont {Cal\'{o}}, \citenamefont {Remirez}, \citenamefont {Chandler}, \citenamefont {Obstein}, \citenamefont {WebsterIII},\ and\ \citenamefont {Valdastri}}]{Campisano2021}%
  \BibitemOpen
  \bibfield  {author} {\bibinfo {author} {\bibfnamefont {F.}~\bibnamefont {Campisano}}, \bibinfo {author} {\bibfnamefont {S.}~\bibnamefont {Cal\'{o}}}, \bibinfo {author} {\bibfnamefont {A.~A.}\ \bibnamefont {Remirez}}, \bibinfo {author} {\bibfnamefont {J.~H.}\ \bibnamefont {Chandler}}, \bibinfo {author} {\bibfnamefont {K.~L.}\ \bibnamefont {Obstein}}, \bibinfo {author} {\bibfnamefont {R.~J.}\ \bibnamefont {WebsterIII}},\ and\ \bibinfo {author} {\bibfnamefont {P.}~\bibnamefont {Valdastri}},\ }\bibfield  {title} {\bibinfo {title} {Closed-loop control of soft continuum manipulators under tip follower actuation},\ }\href@noop {} {\bibfield  {journal} {\bibinfo  {journal} {Int. J. Robot. Res.}\ }\textbf {\bibinfo {volume} {40}},\ \bibinfo {pages} {923} (\bibinfo {year} {2021})}\BibitemShut {NoStop}%
\bibitem [{\citenamefont {Love}(1892)}]{Love1892}%
  \BibitemOpen
  \bibfield  {author} {\bibinfo {author} {\bibfnamefont {A.~E.~H.}\ \bibnamefont {Love}},\ }\href@noop {} {\emph {\bibinfo {title} {A {Treatise} on the {Mathematical} {Theory} of {Elasticity}}}}\ (\bibinfo  {publisher} {Cambridge University Press},\ \bibinfo {address} {Cambridge},\ \bibinfo {year} {1892})\BibitemShut {NoStop}%
\bibitem [{\citenamefont {Ericksen}\ and\ \citenamefont {Truesdell}(1957)}]{Ericksen1957}%
  \BibitemOpen
  \bibfield  {author} {\bibinfo {author} {\bibfnamefont {J.~L.}\ \bibnamefont {Ericksen}}\ and\ \bibinfo {author} {\bibfnamefont {C.}~\bibnamefont {Truesdell}},\ }\bibfield  {title} {\bibinfo {title} {Exact theory of stress and strain in rods and shells},\ }\href@noop {} {\bibfield  {journal} {\bibinfo  {journal} {Arch. Ration. Mech. Anal.}\ }\textbf {\bibinfo {volume} {1}},\ \bibinfo {pages} {295} (\bibinfo {year} {1957})}\BibitemShut {NoStop}%
\bibitem [{\citenamefont {Antman}(2005)}]{Antman2005}%
  \BibitemOpen
  \bibfield  {author} {\bibinfo {author} {\bibfnamefont {S.~S.}\ \bibnamefont {Antman}},\ }\href@noop {} {\emph {\bibinfo {title} {Nonlinear {Problems} of {Elasticity}}}}\ (\bibinfo  {publisher} {Springer},\ \bibinfo {address} {New York},\ \bibinfo {year} {2005})\BibitemShut {NoStop}%
\bibitem [{\citenamefont {Gazzola}\ \emph {et~al.}(2018)\citenamefont {Gazzola}, \citenamefont {Dudte}, \citenamefont {McCormick},\ and\ \citenamefont {Mahadevan}}]{Gazzola2018}%
  \BibitemOpen
  \bibfield  {author} {\bibinfo {author} {\bibfnamefont {M.}~\bibnamefont {Gazzola}}, \bibinfo {author} {\bibfnamefont {L.~H.}\ \bibnamefont {Dudte}}, \bibinfo {author} {\bibfnamefont {A.~G.}\ \bibnamefont {McCormick}},\ and\ \bibinfo {author} {\bibfnamefont {L.}~\bibnamefont {Mahadevan}},\ }\bibfield  {title} {\bibinfo {title} {Forward and inverse problems in the mechanics of soft filaments},\ }\href@noop {} {\bibfield  {journal} {\bibinfo  {journal} {R. Soc. Open Sci.}\ }\textbf {\bibinfo {volume} {5}},\ \bibinfo {pages} {171628} (\bibinfo {year} {2018})}\BibitemShut {NoStop}%
\bibitem [{\citenamefont {Boyer}\ \emph {et~al.}(2021)\citenamefont {Boyer}, \citenamefont {Lebastard}, \citenamefont {Candelier},\ and\ \citenamefont {Renda}}]{Boyer2021}%
  \BibitemOpen
  \bibfield  {author} {\bibinfo {author} {\bibfnamefont {F.}~\bibnamefont {Boyer}}, \bibinfo {author} {\bibfnamefont {V.}~\bibnamefont {Lebastard}}, \bibinfo {author} {\bibfnamefont {F.}~\bibnamefont {Candelier}},\ and\ \bibinfo {author} {\bibfnamefont {F.}~\bibnamefont {Renda}},\ }\bibfield  {title} {\bibinfo {title} {Dynamics of continuum and soft robots: A strain parameterization based approach},\ }\href@noop {} {\bibfield  {journal} {\bibinfo  {journal} {IEEE Trans. Robot.}\ }\textbf {\bibinfo {volume} {37}},\ \bibinfo {pages} {847} (\bibinfo {year} {2021})}\BibitemShut {NoStop}%
\bibitem [{\citenamefont {Naughton}\ \emph {et~al.}(2021)\citenamefont {Naughton}, \citenamefont {Sun}, \citenamefont {Tekinalp}, \citenamefont {Parthasarathy}, \citenamefont {Chowdhary},\ and\ \citenamefont {Gazzola}}]{Gazzola2021}%
  \BibitemOpen
  \bibfield  {author} {\bibinfo {author} {\bibfnamefont {N.}~\bibnamefont {Naughton}}, \bibinfo {author} {\bibfnamefont {J.}~\bibnamefont {Sun}}, \bibinfo {author} {\bibfnamefont {A.}~\bibnamefont {Tekinalp}}, \bibinfo {author} {\bibfnamefont {T.}~\bibnamefont {Parthasarathy}}, \bibinfo {author} {\bibfnamefont {G.}~\bibnamefont {Chowdhary}},\ and\ \bibinfo {author} {\bibfnamefont {M.}~\bibnamefont {Gazzola}},\ }\bibfield  {title} {\bibinfo {title} {Elastica: A compliant mechanics environment for soft robotic control},\ }\href@noop {} {\bibfield  {journal} {\bibinfo  {journal} {IEEE Robot. Autom. Lett.}\ }\textbf {\bibinfo {volume} {6}},\ \bibinfo {pages} {3389} (\bibinfo {year} {2021})}\BibitemShut {NoStop}%
\bibitem [{\citenamefont {Chen}\ \emph {et~al.}(2022)\citenamefont {Chen}, \citenamefont {Liu},\ and\ \citenamefont {Han}}]{Chen2022}%
  \BibitemOpen
  \bibfield  {author} {\bibinfo {author} {\bibfnamefont {Z.}~\bibnamefont {Chen}}, \bibinfo {author} {\bibfnamefont {Z.}~\bibnamefont {Liu}},\ and\ \bibinfo {author} {\bibfnamefont {X.}~\bibnamefont {Han}},\ }\bibfield  {title} {\bibinfo {title} {Model analysis of robotic soft arms including external force effects},\ }\href@noop {} {\bibfield  {journal} {\bibinfo  {journal} {Micromachines}\ }\textbf {\bibinfo {volume} {13}},\ \bibinfo {pages} {1486} (\bibinfo {year} {2022})}\BibitemShut {NoStop}%
\bibitem [{\citenamefont {Tekinalp}\ \emph {et~al.}(2025)\citenamefont {Tekinalp}, \citenamefont {Bhosale}, \citenamefont {Cui}, \citenamefont {Chan},\ and\ \citenamefont {Gazzola}}]{Gazzola2025}%
  \BibitemOpen
  \bibfield  {author} {\bibinfo {author} {\bibfnamefont {A.}~\bibnamefont {Tekinalp}}, \bibinfo {author} {\bibfnamefont {Y.}~\bibnamefont {Bhosale}}, \bibinfo {author} {\bibfnamefont {S.}~\bibnamefont {Cui}}, \bibinfo {author} {\bibfnamefont {F.~K.}\ \bibnamefont {Chan}},\ and\ \bibinfo {author} {\bibfnamefont {M.}~\bibnamefont {Gazzola}},\ }\bibfield  {title} {\bibinfo {title} {Self-propelling, soft, and slender structures in fluids: {C}osserat rods immersed in the velocity--vorticity formulation of the incompressible {N}avier--{S}tokes equations},\ }\href@noop {} {\bibfield  {journal} {\bibinfo  {journal} {Comput. Methods Appl. Mech. Eng.}\ }\textbf {\bibinfo {volume} {440}},\ \bibinfo {pages} {117910} (\bibinfo {year} {2025})}\BibitemShut {NoStop}%
\bibitem [{\citenamefont {Laskar}\ and\ \citenamefont {Adhikari}(2017)}]{Laskar2017}%
  \BibitemOpen
  \bibfield  {author} {\bibinfo {author} {\bibfnamefont {A.}~\bibnamefont {Laskar}}\ and\ \bibinfo {author} {\bibfnamefont {R.}~\bibnamefont {Adhikari}},\ }\bibfield  {title} {\bibinfo {title} {Filament actuation by an active colloid at low {R}eynolds number},\ }\href@noop {} {\bibfield  {journal} {\bibinfo  {journal} {New J. Phys.}\ }\textbf {\bibinfo {volume} {19}},\ \bibinfo {pages} {033021} (\bibinfo {year} {2017})}\BibitemShut {NoStop}%
\bibitem [{\citenamefont {De~Canio}\ \emph {et~al.}(2017)\citenamefont {De~Canio}, \citenamefont {Lauga},\ and\ \citenamefont {Goldstein}}]{DeCanio2017}%
  \BibitemOpen
  \bibfield  {author} {\bibinfo {author} {\bibfnamefont {G.}~\bibnamefont {De~Canio}}, \bibinfo {author} {\bibfnamefont {E.}~\bibnamefont {Lauga}},\ and\ \bibinfo {author} {\bibfnamefont {R.~E.}\ \bibnamefont {Goldstein}},\ }\bibfield  {title} {\bibinfo {title} {Spontaneous oscillations of elastic filaments induced by molecular motors},\ }\href@noop {} {\bibfield  {journal} {\bibinfo  {journal} {J. R. Soc. Interface}\ }\textbf {\bibinfo {volume} {14}},\ \bibinfo {pages} {20170491} (\bibinfo {year} {2017})}\BibitemShut {NoStop}%
\bibitem [{\citenamefont {Ling}\ \emph {et~al.}(2018)\citenamefont {Ling}, \citenamefont {Guo},\ and\ \citenamefont {Kanso}}]{Ling2018}%
  \BibitemOpen
  \bibfield  {author} {\bibinfo {author} {\bibfnamefont {F.}~\bibnamefont {Ling}}, \bibinfo {author} {\bibfnamefont {H.}~\bibnamefont {Guo}},\ and\ \bibinfo {author} {\bibfnamefont {E.}~\bibnamefont {Kanso}},\ }\bibfield  {title} {\bibinfo {title} {Instability-driven oscillations of elastic microfilaments},\ }\href@noop {} {\bibfield  {journal} {\bibinfo  {journal} {J. R. Soc. Interface}\ }\textbf {\bibinfo {volume} {15}},\ \bibinfo {pages} {20180594} (\bibinfo {year} {2018})}\BibitemShut {NoStop}%
\bibitem [{\citenamefont {Man}\ and\ \citenamefont {Kanso}(2019)}]{Man2019}%
  \BibitemOpen
  \bibfield  {author} {\bibinfo {author} {\bibfnamefont {Y.}~\bibnamefont {Man}}\ and\ \bibinfo {author} {\bibfnamefont {E.}~\bibnamefont {Kanso}},\ }\bibfield  {title} {\bibinfo {title} {Morphological transitions of axially-driven microfilaments},\ }\href@noop {} {\bibfield  {journal} {\bibinfo  {journal} {Soft Matter}\ }\textbf {\bibinfo {volume} {15}},\ \bibinfo {pages} {5163} (\bibinfo {year} {2019})}\BibitemShut {NoStop}%
\bibitem [{\citenamefont {Fily}\ \emph {et~al.}(2020)\citenamefont {Fily}, \citenamefont {Subramanian}, \citenamefont {Schneider}, \citenamefont {Chelakkot},\ and\ \citenamefont {Gopinath}}]{Fily2020}%
  \BibitemOpen
  \bibfield  {author} {\bibinfo {author} {\bibfnamefont {Y.}~\bibnamefont {Fily}}, \bibinfo {author} {\bibfnamefont {P.}~\bibnamefont {Subramanian}}, \bibinfo {author} {\bibfnamefont {T.~M.}\ \bibnamefont {Schneider}}, \bibinfo {author} {\bibfnamefont {R.}~\bibnamefont {Chelakkot}},\ and\ \bibinfo {author} {\bibfnamefont {A.}~\bibnamefont {Gopinath}},\ }\bibfield  {title} {\bibinfo {title} {Buckling instabilities and spatio-temporal dynamics of active elastic filaments},\ }\href@noop {} {\bibfield  {journal} {\bibinfo  {journal} {J. R. Soc. Interface}\ }\textbf {\bibinfo {volume} {17}},\ \bibinfo {pages} {20190794} (\bibinfo {year} {2020})}\BibitemShut {NoStop}%
\bibitem [{\citenamefont {Clarke}\ \emph {et~al.}(2024)\citenamefont {Clarke}, \citenamefont {Hwang},\ and\ \citenamefont {Keaveny}}]{Clarke2024}%
  \BibitemOpen
  \bibfield  {author} {\bibinfo {author} {\bibfnamefont {B.}~\bibnamefont {Clarke}}, \bibinfo {author} {\bibfnamefont {Y.}~\bibnamefont {Hwang}},\ and\ \bibinfo {author} {\bibfnamefont {E.~E.}\ \bibnamefont {Keaveny}},\ }\bibfield  {title} {\bibinfo {title} {Bifurcations and nonlinear dynamics of the follower force model for active filaments},\ }\href@noop {} {\bibfield  {journal} {\bibinfo  {journal} {Phys. Rev. Fluids}\ }\textbf {\bibinfo {volume} {9}},\ \bibinfo {pages} {073101} (\bibinfo {year} {2024})}\BibitemShut {NoStop}%
\bibitem [{\citenamefont {Link}\ \emph {et~al.}(2024)\citenamefont {Link}, \citenamefont {Guy}, \citenamefont {Thomases},\ and\ \citenamefont {Arratia}}]{Link2024}%
  \BibitemOpen
  \bibfield  {author} {\bibinfo {author} {\bibfnamefont {K.~G.}\ \bibnamefont {Link}}, \bibinfo {author} {\bibfnamefont {R.~D.}\ \bibnamefont {Guy}}, \bibinfo {author} {\bibfnamefont {B.}~\bibnamefont {Thomases}},\ and\ \bibinfo {author} {\bibfnamefont {P.~E.}\ \bibnamefont {Arratia}},\ }\bibfield  {title} {\bibinfo {title} {Effect of fluid elasticity on the emergence of oscillations in an active elastic filament},\ }\href@noop {} {\bibfield  {journal} {\bibinfo  {journal} {J. R. Soc. Interface}\ }\textbf {\bibinfo {volume} {21}},\ \bibinfo {pages} {20240046} (\bibinfo {year} {2024})}\BibitemShut {NoStop}%
\bibitem [{\citenamefont {Schnitzer}(2025)}]{Schnitzer2025}%
  \BibitemOpen
  \bibfield  {author} {\bibinfo {author} {\bibfnamefont {O.}~\bibnamefont {Schnitzer}},\ }\bibfield  {title} {\bibinfo {title} {Onset of spontaneous beating and whirling in the follower force model of an active filament},\ }\href@noop {} {\bibfield  {journal} {\bibinfo  {journal} {J. Fluid Mech.}\ }\textbf {\bibinfo {volume} {1007}},\ \bibinfo {pages} {A65} (\bibinfo {year} {2025})}\BibitemShut {NoStop}%
\bibitem [{\citenamefont {Beck}(1952)}]{Beck1952}%
  \BibitemOpen
  \bibfield  {author} {\bibinfo {author} {\bibfnamefont {M.}~\bibnamefont {Beck}},\ }\bibfield  {title} {\bibinfo {title} {Die knicklast des einseitig eingespannten, tangential gedr\"{u}ckten stabes},\ }\href@noop {} {\bibfield  {journal} {\bibinfo  {journal} {Z. Angew. Math. Phys.}\ }\textbf {\bibinfo {volume} {3}},\ \bibinfo {pages} {225} (\bibinfo {year} {1952})}\BibitemShut {NoStop}%
\bibitem [{\citenamefont {Bolotin}(1963)}]{Bolotin1963}%
  \BibitemOpen
  \bibfield  {author} {\bibinfo {author} {\bibfnamefont {V.~V.}\ \bibnamefont {Bolotin}},\ }\href@noop {} {\emph {\bibinfo {title} {Nonconservative {Problems} of the {Theory} of {Elastic} {Stability}}}},\ edited by\ \bibinfo {editor} {\bibfnamefont {G.}~\bibnamefont {Herrmann}}\ (\bibinfo  {publisher} {Pergamon Press},\ \bibinfo {address} {London},\ \bibinfo {year} {1963})\BibitemShut {NoStop}%
\bibitem [{\citenamefont {Ziegler}(1968)}]{Ziegler1968}%
  \BibitemOpen
  \bibfield  {author} {\bibinfo {author} {\bibfnamefont {H.}~\bibnamefont {Ziegler}},\ }\href@noop {} {\emph {\bibinfo {title} {Principles of {Structural} {Stability}}}}\ (\bibinfo  {publisher} {Blaisdell Publishing Company},\ \bibinfo {address} {Waltham, MA},\ \bibinfo {year} {1968})\BibitemShut {NoStop}%
\bibitem [{\citenamefont {Carr}\ and\ \citenamefont {Malhardeen}(1979)}]{Carr1979}%
  \BibitemOpen
  \bibfield  {author} {\bibinfo {author} {\bibfnamefont {J.}~\bibnamefont {Carr}}\ and\ \bibinfo {author} {\bibfnamefont {M.~Z.~M.}\ \bibnamefont {Malhardeen}},\ }\bibfield  {title} {\bibinfo {title} {Beck's problem},\ }\href@noop {} {\bibfield  {journal} {\bibinfo  {journal} {SIAM J. Appl. Math.}\ }\textbf {\bibinfo {volume} {37}},\ \bibinfo {pages} {261} (\bibinfo {year} {1979})}\BibitemShut {NoStop}%
\bibitem [{\citenamefont {Chen}(1987)}]{Chen1987}%
  \BibitemOpen
  \bibfield  {author} {\bibinfo {author} {\bibfnamefont {M.}~\bibnamefont {Chen}},\ }\bibfield  {title} {\bibinfo {title} {Hopf bifurcation in {B}eck's problem},\ }\href@noop {} {\bibfield  {journal} {\bibinfo  {journal} {Nonlinear Analysis: Theory, Methods \& Applications}\ }\textbf {\bibinfo {volume} {11}},\ \bibinfo {pages} {1061} (\bibinfo {year} {1987})}\BibitemShut {NoStop}%
\bibitem [{\citenamefont {Koch}\ and\ \citenamefont {Antman}(2000)}]{Koch2000}%
  \BibitemOpen
  \bibfield  {author} {\bibinfo {author} {\bibfnamefont {H.}~\bibnamefont {Koch}}\ and\ \bibinfo {author} {\bibfnamefont {S.~S.}\ \bibnamefont {Antman}},\ }\bibfield  {title} {\bibinfo {title} {Stability and {H}opf bifurcation for fully nonlinear parabolic-hyperbolic equations},\ }\href@noop {} {\bibfield  {journal} {\bibinfo  {journal} {SIAM J. Math. Anal.}\ }\textbf {\bibinfo {volume} {32}},\ \bibinfo {pages} {360} (\bibinfo {year} {2000})}\BibitemShut {NoStop}%
\bibitem [{\citenamefont {Wang}(2004)}]{Wang2004}%
  \BibitemOpen
  \bibfield  {author} {\bibinfo {author} {\bibfnamefont {Q.}~\bibnamefont {Wang}},\ }\bibfield  {title} {\bibinfo {title} {A comprehensive stability analysis of a cracked beam subjected to follower compression},\ }\href@noop {} {\bibfield  {journal} {\bibinfo  {journal} {Int. J. Solids Struct.}\ }\textbf {\bibinfo {volume} {41}},\ \bibinfo {pages} {4875} (\bibinfo {year} {2004})}\BibitemShut {NoStop}%
\bibitem [{\citenamefont {Yan}\ \emph {et~al.}(2025)\citenamefont {Yan}, \citenamefont {Warda}, \citenamefont {N{\'e}meth}, \citenamefont {Kikuchi},\ and\ \citenamefont {Adhikari}}]{Yan2025}%
  \BibitemOpen
  \bibfield  {author} {\bibinfo {author} {\bibfnamefont {M.}~\bibnamefont {Yan}}, \bibinfo {author} {\bibfnamefont {M.}~\bibnamefont {Warda}}, \bibinfo {author} {\bibfnamefont {B.}~\bibnamefont {N{\'e}meth}}, \bibinfo {author} {\bibfnamefont {L.}~\bibnamefont {Kikuchi}},\ and\ \bibinfo {author} {\bibfnamefont {R.}~\bibnamefont {Adhikari}},\ }\href {https://arxiv.org/abs/2510.18097} {\bibinfo {title} {Geometric field theory for elastohydrodynamics of {C}osserat rods}} (\bibinfo {year} {2025}),\ \Eprint {https://arxiv.org/abs/2510.18097} {arXiv:2510.18097 [cond-mat.soft]} \BibitemShut {NoStop}%
\bibitem [{\citenamefont {Garg}\ and\ \citenamefont {Kumar}(2023)}]{Garg2023}%
  \BibitemOpen
  \bibfield  {author} {\bibinfo {author} {\bibfnamefont {M.}~\bibnamefont {Garg}}\ and\ \bibinfo {author} {\bibfnamefont {A.}~\bibnamefont {Kumar}},\ }\bibfield  {title} {\bibinfo {title} {A slender body theory for the motion of special {C}osserat filaments in {S}tokes flow},\ }\href@noop {} {\bibfield  {journal} {\bibinfo  {journal} {Mathematics and Mechanics of Solids}\ }\textbf {\bibinfo {volume} {28}},\ \bibinfo {pages} {692} (\bibinfo {year} {2023})}\BibitemShut {NoStop}%
\bibitem [{\citenamefont {Kikuchi}(2023)}]{KikuchiThesis}%
  \BibitemOpen
  \bibfield  {author} {\bibinfo {author} {\bibfnamefont {T.}~\bibnamefont {Kikuchi}},\ }\emph {\bibinfo {title} {Rare events and dynamics in non-equilibrium systems}},\ \href {https://doi.org/10.17863/CAM.105378} {Ph.D. thesis},\ \bibinfo  {school} {Apollo - University of Cambridge Repository} (\bibinfo {year} {2023})\BibitemShut {NoStop}%
\bibitem [{\citenamefont {Kikuchi}\ and\ \citenamefont {Adhikari}(2023)}]{Kikuchi2023}%
  \BibitemOpen
  \bibfield  {author} {\bibinfo {author} {\bibfnamefont {L.}~\bibnamefont {Kikuchi}}\ and\ \bibinfo {author} {\bibfnamefont {R.}~\bibnamefont {Adhikari}},\ }\href@noop {} {\bibinfo {title} {Cartan media: geometric continuum mechanics in homogeneous spaces}} (\bibinfo {year} {2023}),\ \Eprint {https://arxiv.org/abs/2310.01388} {arXiv:2310.01388 [cond-mat.soft]} \BibitemShut {NoStop}%
\bibitem [{\citenamefont {Ashida}\ \emph {et~al.}(2020)\citenamefont {Ashida}, \citenamefont {Gong},\ and\ \citenamefont {Ueda}}]{Ashida2020}%
  \BibitemOpen
  \bibfield  {author} {\bibinfo {author} {\bibfnamefont {Y.}~\bibnamefont {Ashida}}, \bibinfo {author} {\bibfnamefont {Z.}~\bibnamefont {Gong}},\ and\ \bibinfo {author} {\bibfnamefont {M.}~\bibnamefont {Ueda}},\ }\bibfield  {title} {\bibinfo {title} {Non-hermitian physics},\ }\href@noop {} {\bibfield  {journal} {\bibinfo  {journal} {Advances in Physics}\ }\textbf {\bibinfo {volume} {69}},\ \bibinfo {pages} {249} (\bibinfo {year} {2020})}\BibitemShut {NoStop}%
\bibitem [{\citenamefont {Jastrzebski}(1959)}]{Jastrzebski1959}%
  \BibitemOpen
  \bibfield  {author} {\bibinfo {author} {\bibfnamefont {Z.}~\bibnamefont {Jastrzebski}},\ }\href@noop {} {\emph {\bibinfo {title} {Nature and Properties of Engineering Materials}}},\ Wiley international edition\ (\bibinfo  {publisher} {Wiley},\ \bibinfo {year} {1959})\BibitemShut {NoStop}%
\bibitem [{imp()}]{implementation_note}%
  \BibitemOpen
  \href@noop {} {}\bibinfo {note} {Mathematica code and implementation details are available from the author upon request.}\BibitemShut {Stop}%
\bibitem [{\citenamefont {Krishnaprasad}\ and\ \citenamefont {Tsakiris}(1994)}]{Krishnaprasad1994}%
  \BibitemOpen
  \bibfield  {author} {\bibinfo {author} {\bibfnamefont {P.~S.}\ \bibnamefont {Krishnaprasad}}\ and\ \bibinfo {author} {\bibfnamefont {D.~P.}\ \bibnamefont {Tsakiris}},\ }\bibfield  {title} {\bibinfo {title} {G-snakes: Nonholonomic kinematic chains on lie groups},\ }in\ \href@noop {} {\emph {\bibinfo {booktitle} {Proceedings of the 33rd IEEE Conference on Decision and Control}}}\ (\bibinfo  {publisher} {IEEE},\ \bibinfo {address} {Lake Buena Vista, FL},\ \bibinfo {year} {1994})\ pp.\ \bibinfo {pages} {2955--2960}\BibitemShut {NoStop}%
\bibitem [{\citenamefont {Holm}\ \emph {et~al.}(2012)\citenamefont {Holm}, \citenamefont {Ivanov},\ and\ \citenamefont {Percival}}]{Holm2012}%
  \BibitemOpen
  \bibfield  {author} {\bibinfo {author} {\bibfnamefont {D.}~\bibnamefont {Holm}}, \bibinfo {author} {\bibfnamefont {R.}~\bibnamefont {Ivanov}},\ and\ \bibinfo {author} {\bibfnamefont {J.~R.}\ \bibnamefont {Percival}},\ }\bibfield  {title} {\bibinfo {title} {G-strands},\ }\href@noop {} {\bibfield  {journal} {\bibinfo  {journal} {J. Nonlinear Sci.}\ }\textbf {\bibinfo {volume} {22}},\ \bibinfo {pages} {517} (\bibinfo {year} {2012})}\BibitemShut {NoStop}%
\end{thebibliography}%

\end{document}